\documentclass{article}

\usepackage{amsmath,amssymb}
\usepackage{array}
\usepackage{arxiv}
\usepackage{booktabs}
\usepackage[labelfont=bf]{caption}
\usepackage{cite}
\usepackage{float}
\usepackage{graphicx}
\usepackage{hyperref}
\usepackage[framemethod=tikz]{mdframed}
\usepackage{xcolor}

\newif\ifcomment
\commenttrue    

\newif\iffigabbrv
\figabbrvfalse   
\newcommand{\figtext}{\iffigabbrv Fig.\else Figure\fi}
\newcommand{\figstext}{\iffigabbrv Figs.\else Figures\fi}

\newcommand{\boxtext}[1]{\begin{mdframed}[roundcorner=5pt]#1\end{mdframed}}

\title{Preventing Extreme Polarization of Political Attitudes}
\author{
    Robert Axelrod \\
        School of Public Policy \\
        University of Michigan \\
        Ann Arbor, MI 48109, USA \\
        \texttt{axe@umich.edu} \\
    \And
    Joshua J.\ Daymude \\
        Biodesign Institute \\
        Arizona State University \\
        Tempe, AZ 85281, USA \\
        \texttt{jdaymude@asu.edu} \\
    \And
    Stephanie Forrest \\
        Biodesign Institute and CIDSE \\
        Arizona State University \\
        Tempe, AZ 85281, USA \\
        \texttt{steph@asu.edu} \\
}
\date{}

\begin{document}

\maketitle

\begin{abstract}
    Extreme polarization can undermine democracy by making compromise impossible and transforming politics into a zero-sum game.
    \textit{Ideological polarization}---the extent to which political views are widely dispersed---is already strong among elites, but less so among the general public (McCarty, 2019, p.\ 50--68).
    Strong mutual distrust and hostility between Democrats and Republicans in the U.S., combined with the elites' already strong ideological polarization, could lead to increasing ideological polarization among the public.
    The paper addresses two questions: (1) Is there a level of ideological polarization above which polarization feeds upon itself to become a runaway process? (2) If so, what policy interventions could prevent such dangerous positive feedback loops?
    To explore these questions, we present an agent-based model of ideological polarization that differentiates between the tendency for two actors to interact (\textit{exposure}) and how they respond when interactions occur, positing that interaction between similar actors \textit{reduces} their difference while interaction between dissimilar actors \textit{increases} their difference.
    Our analysis explores the effects on polarization of different levels of tolerance to other views, responsiveness to other views, exposure to dissimilar actors, multiple ideological dimensions, economic self-interest, and external shocks.
    The results suggest strategies for preventing, or at least slowing, the development of extreme polarization.
\end{abstract}

\section*{Introduction}

Extreme polarization can undermine democracy by making compromise impossible~\cite{Mason2016-crosscuttingcalm,Patty2019-moderatesextremists} and transforming politics into a zero-sum game, as James Madison observed in Federalist No.\ 10~\cite{Madison1787-federalist10}.
When this occurs, even a democratically elected majority may seek to solidify its control over political power by weakening the institutions and norms that ordinarily support the turnover of elites.
There are many motivating examples of this dysfunctional political polarization: e.g., the rise of Hitler, the American Civil War, the destruction of democracy in Venezuela, the increasing threats to democracy in Hungary, and the growing animosities in American politics over the past two or three decades~\cite{Levitsky2018-democraciesdie}.

At least two kinds of polarization, if carried to extremes, can undermine democracy.
\textit{Affective polarization} is already a serious problem in the U.S.; e.g., Americans increasingly dislike and distrust those of the other party, whether Democrat or Republican~\cite{Iyengar2019-affective,Mason2016-crosscuttingcalm}.
The other kind, \textit{ideological polarization}, is the extent to which political views are widely dispersed.
Ideological polarization is already strong among elites but is less pronounced among the general public; see~\cite{Yang2020-partypolarization} and p.\ 50--68 of~\cite{McCarty2019-polarization}.
In the future, ideological polarization among the U.S.\ public may increase due to the already strong affective polarization, rising social inequality, and the collapse of cross-cutting belief structures into consolidated clusters~\cite{DellaPosta2020-oilspillmodel,Putnam2020-upswingtogether}.
Therefore, it is important to understand how to prevent the public from reaching dangerous degrees of ideological polarization.

This paper focuses on ideological polarization (henceforth ``polarization'') among the general public and asks:
Is there a level of ideological polarization above which polarization becomes a runaway process?
And, if so, what policy interventions can prevent such a dangerous positive feedback loop?

To address these questions, we developed an agent-based model (ABM) of ideological polarization to explore situations in which many actors influence each other in ways that don't lend themselves to equation-based models.
In the ABM paradigm, each individual actor is represented explicitly and rules specify the mechanisms for interaction between actors.
Simple ABMs, such as ours, are not intended to predict any particular historical event or future outcome.
Instead, they can provide insight about important mechanisms and the role they play in determining system trajectories, e.g., to highlight the consequences of a few simple assumptions about how people are influenced by each other.
By design, ABMs can capture a distribution of possible outcomes, characterizing both typical and rare behavior.

To study ideological polarization, we selected a small set of mechanisms that influence opinion change: attraction to those with similar ideological (i.e., political) positions and repulsion from those with dissimilar positions.
In our \textit{Attraction-Repulsion Model (ARM)}, the actors are assumed to follow a few simple rules about giving and receiving influence.
The rules are not based on principles of rational calculation, i.e., costs and benefits, or the forward-looking strategic analysis typical of game theory.
Instead, the actors simply adapt their position in ideological space based on interactions with other actors.
Note that when one actor changes its position, the environment of all the other actors is affected.
Based on idealized simple rules of interaction, we investigate the emergent properties of the system over time.
Because the proposed mechanisms can exist alongside other mechanisms, they are complementary with other treatments of polarization.

Most agent-based~\cite{Flache2017-modelssocialinfluence,Flache2018-monoculturediversity} and statistical physics~\cite{Castellano2009-statphyssocial,Redner2019-realvotermodels} models of opinion change, including Axelrod's culture model~\cite{Axelrod1997-culturemodel}, investigate the effects of \textit{homophily} and \textit{assimilation}: the tendency to interact with and attract towards others with similar opinions.
Our ARM joins a growing body of literature that additionally considers \textit{differentiation} (or \textit{negative influence}): the tendency to amplify difference from others with dissimilar opinions~\cite{Baldassarri2007-dynamicspolarization,VazMartins2010-massmedia,Flache2011-smallworlds,Krause2019-repulsiondebate,Santos2021-socialmatching,Szymanski2021-tippingpoints}.
Although empirical evidence for differentiation is mixed---e.g., negative interactions are apparent in the U.S.\ Senate~\cite{Liu2015-pullingcloser} and on social media~\cite{Bail2018-socialmediapolarization} but are not always observed in group-discussion and opinion-exchange experiments~\cite{Mas2013-diffwithoutdistance,Takacs2016-discrepancydisliking}---differentiation can capture the empirical effects of external messaging~\cite{VazMartins2010-massmedia} and public debate on controversial topics~\cite{Krause2019-repulsiondebate} that purely assimilative models do not.
Unlike most other models with both assimilation and differentiation~\cite{VazMartins2010-massmedia,Flache2011-smallworlds,Krause2019-repulsiondebate,Santos2021-socialmatching,Szymanski2021-tippingpoints}, our model assumes that (\textit{i}) both the likelihood of interaction and the magnitude and direction of opinion change are affected by ideological distance and (\textit{ii}) these effects are uncorrelated.
A notable exception is the model of Baldassarri and Bearman~\cite{Baldassarri2007-dynamicspolarization}, but their model includes several additional features such as issue engagement and perception, complicating the task of characterizing the effects of interaction and opinion change.

Our analysis of polarization has several novel features.
First, our model of polarization dynamics is unique in its simplicity, representing opinion change based only on an individual's attraction or repulsion to others' positions.
Its very simplicity allows us to attribute the exact mechanisms that yield particular outcomes.
Second, unlike many models of opinion change, the ARM avoids the common assumption that the direction and magnitude of opinion change are correlated which may not necessarily be true.
Third, in the context of models of polarization dynamics that include repulsion as well as attraction, our discoveries include:
\begin{enumerate}
    \item The identification of conditions under which a population first approaches convergence around a moderate position but then reverses direction and becomes highly polarized, i.e., conditions under which the center does not hold.

    \item The identification of conditions under which a few extremists can actually help prevent polarization.

    \item The discovery that even weak attraction to one's own initial position (such as the effect of economic self-interest) can prevent polarization.

    \item Insight into the seeming paradox that---contrary to many policy proposals---exposure to dissimilar views can actually exasperate rather than alleviate polarization.
\end{enumerate}

\section*{The Attraction-Repulsion Model (ARM)}

The ARM has only two rules.
Stated informally, one rule says that an actor tends to interact with those who have similar views.
The second rule says that a (pairwise) interaction between similar actors reduces their difference, while an interaction between dissimilar actors increases their difference.
The ARM is, in fact, a Markov process where the choice of interacting actors is stochastic but interactions' effects are deterministic.
Conditioned on a given sequence of interacting pairs of actors, the ARM is a deterministic dynamical system.
While some analytic results from Markov theory are potentially available, a simulation approach is preferred here.

In particular, we assume a population of $N$ actors whose ideological positions are distributed in $D$-dimensional space, where each dimension corresponds to an ideological, or political, issue.
Each actor has a position between 0.0 and 1.0 on any given dimension;\footnote{We assume that ideological positions are bounded to prevent repulsion from causing unbounded moves and because ideological positions are typically measured in surveys by questions with a limited integer range, e.g., on a five or seven point scale.} our results focus primarily on one-dimensional space.
Many ABMs assume that actors' positions are initially distributed uniformly or randomly; we instead assume the normal (Gaussian) distribution which is more realistic for modeling human political views~\cite{Yang2020-partypolarization}.\footnote{Populations initialized according to estimates of one-dimensional latent ideology extracted from 2020 survey data~\cite{Schaffner2021-cces2020,Tausanovitch2013-measuringpolicy} yield similar behavior (see the Supplementary Materials and \figtext~\ref{fig:empiricalinit}).}
Initial positions are normally distributed with a mean of $0.5$ and standard deviation of $\sigma = 0.2$, unless stated otherwise.
The two rules are defined as follows:

\textbf{Interaction Rule.} At each step, a random actor selects another actor uniformly at random and they interact with probability $(1/2)^{d/E}$, where $d$ is the distance between them and $E$ is a model parameter capturing actors' \textit{exposure} to other points of view.

This rule reflects the idea that the probability of interaction decays with the distance between ideological positions, and in our model the decay is exponential.
The scaling factor is the ``halving distance'' of exposure $E$.
As an example, if two actors are distance $E$ apart, they have a $50\%$ chance of interacting; if they are $2E$ apart, they have a $25\%$ chance; and if they are $E/2$ apart, they have a $1/\sqrt{2} \approx 71\%$ chance.
A large exposure means that an actor is almost as likely to interact with someone with whom it disagrees as with someone who has a similar position, i.e., the population is largely unsorted.

\textbf{Attraction-Repulsion (AR) Rule.} All actors have identical \textit{tolerance}, $T$, that determines whether the result of an interaction is attractive or repulsive.
If an actor tolerates the position of the other, the actor moves a fraction $R$ (for \textit{responsiveness}) of the distance toward the other.
Otherwise, the actor moves the same distance away from the other, subject to the boundaries.\footnote{Instead of deterministic attraction within a tolerance distance and repulsion beyond it, one could model the probability of repulsion as a smooth stochastic function that increases with actors' pairwise distance to obtain similar results (see the Supplementary Materials).}

A motivating example for the AR rule is the Bail et al.~\cite{Bail2018-socialmediapolarization} study showing that exposure to opposing views on social media can lead to repulsion.
As an example of AR, if an actor is at $0.4$ and $T = 0.15$, the actor will move closer to anyone between $0.25$ and $0.55$.
If $R = 0.25$ and the actor interacts with another at $0.5$, the actor will move a quarter of the way from its position at $0.4$ to the other's position at $0.5$, resulting in a new position of $0.425$.
On the other hand, if the other actor is outside of the tolerance range, the actor will be repulsed.
For example, if the other actor were at $0.1$, the distance between them would be $0.4 - 0.1 = 0.3$ which is greater than the tolerance of $0.15$.
In that case, the actor would move a fraction $R$ of the distance between them away from the other, resulting in a position of $0.4 + (0.25)(0.3) = 0.475$.

Unless the actor is already near an extreme, an interesting effect of this rule is that repulsion moves actors further than attraction does.
The distance moved is always an $R$-fraction of the distance between the interacting actors which is greater than $T$ for repulsion and less than $T$ for attraction.
Yet actors within distance $T$ are also more likely to interact because they are closer.
\textit{Repulsion is less likely than attraction, but leads to greater movement when it does occur.}

We operationalize the degree of polarization as the \textit{variance} of political positions in a population.
For example, suppose there is one ideological dimension with actors distributed according to a normal (Gaussian) distribution with standard deviation $\sigma = 0.2$.
Since variance is $\sigma^2$, the polarization is $\sigma^2 = 0.04$.
Each ideological dimension ranges from 0.0 to 1.0, so the maximum polarization for the one-dimensional setting is $0.25$ when half the population are extremists at zero and the other half are at one.
The minimum polarization is zero, which occurs if the population converges to a single point.

\begin{table}[tbh]
    \centering
    \caption{Default parameter values for the Attraction-Repulsion Model.}
    \label{tab:parameters}
    \begin{tabular}{m{10cm}l}
        Description & Default Value \\
        \midrule
        Number of actors & $N = 100$ \\
        \smallskip
        Number of ideological dimensions & $D = 1$ \\
        \smallskip
        Mean and standard deviation of actors' initial normal distribution along an ideological dimension & $(0.5, 0.2)$ \\
        \smallskip
        Exposure, the degree to which actors interact with dissimilar points of view expressed as the halving distance & $E = 0.1$ \\
        \smallskip
        Tolerance, the distance within which interactions are attractive and beyond which interactions are repulsive & $T = 0.25$ \\
        \smallskip
        Responsiveness, the fractional distance an actor's ideological position moves as a result of an interaction & $R = 0.25$ \\
        \bottomrule
    \end{tabular}
\end{table}

Table~\ref{tab:parameters} gives the default parameter values used for our experiments.
There is no natural calibration of time in the model, so we adopt the unit of actor activations, called steps.
A helpful way to think about time is to suppose that on average each actor has one activation per day.
Of course, only some of the activations result in interaction and movement, with similar actors more likely than distant actors to interact.
With 100 actors, 36,500 steps of the simulation represents one year, and one million steps represents about 27 years.

\section*{Results}

The findings of our simulation experiments are presented in six categories: tolerance, responsiveness, exposure, multiple ideological dimensions, economic self-interest, and external shock.
See the Supplementary Materials for details on the model implementation.

\subsection*{Tolerance}

\begin{figure}[tbh]
    \centering
    \includegraphics[width=0.6\linewidth]{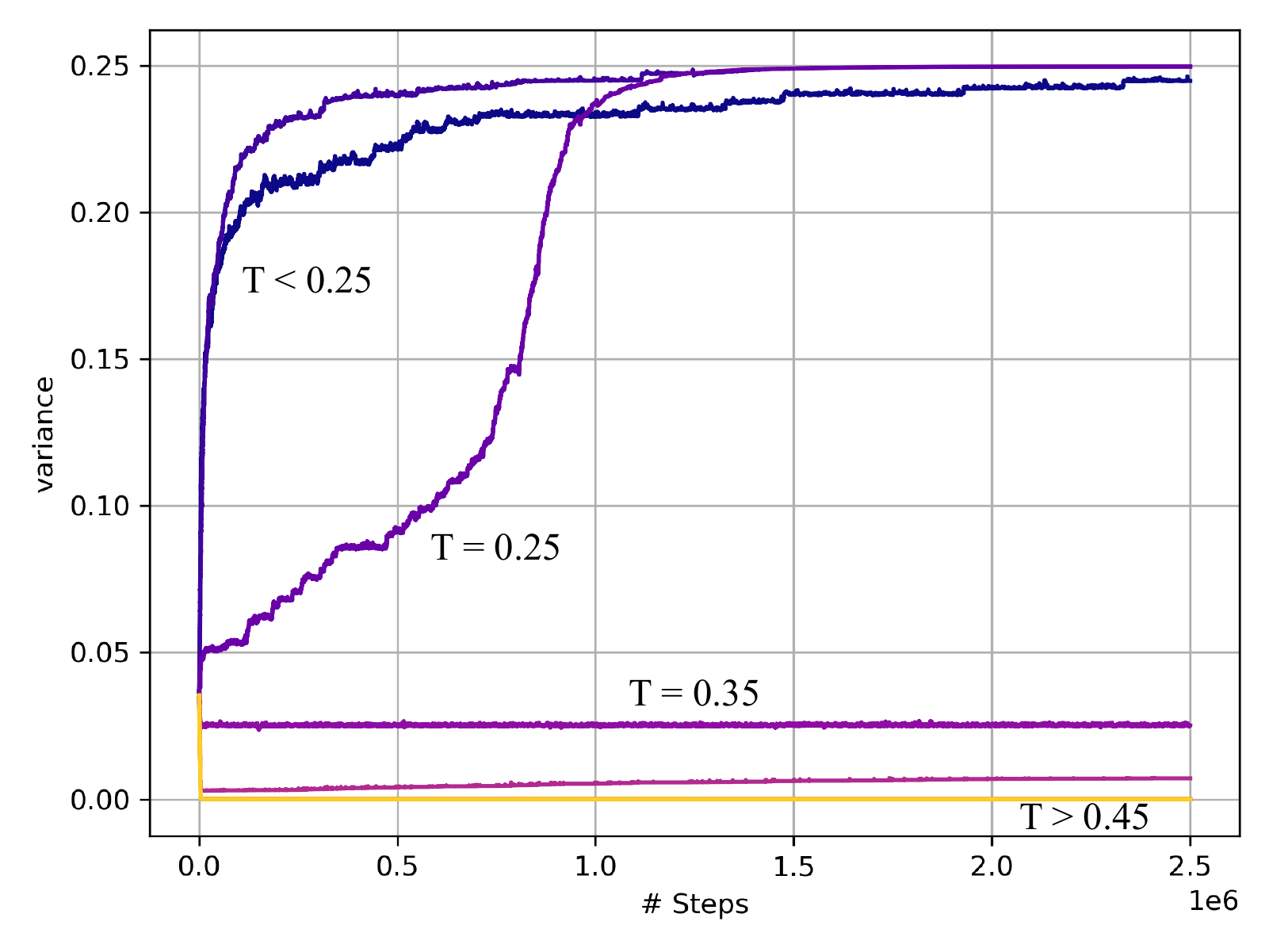}
    \caption{\textbf{The Effects of Tolerance (T).}
    Polarization of the population's ideological positions over time when varying tolerance over the range $T = 0.05, 0.15, \ldots, 0.95$ (dark blue to yellow).
    Low tolerance ($T \leq 0.25$) leads to extreme polarization, intermediate tolerance ($0.35 \leq  T \leq 0.45$) leads to small but non-zero polarization, and high tolerance ($T \geq 0.55$) leads to convergence.}
    \label{fig:toleranceevo}
\end{figure}

We begin by exploring the effects of \textit{tolerance $(T)$}, i.e., the level of ideological difference that actors find attractive rather than repulsive.
\figtext~\ref{fig:toleranceevo} shows the development of polarization over time for different values of $T$.
As expected, when actors have low $T$ ($\leq 0.15$), many of their interactions are outside their tolerance, leading to repulsion which increases their distance.
This produces extreme polarization with roughly half the population at one extreme and the other half at the other.
It is also unsurprising that when $T$ is sufficiently high ($\geq 0.55$), interactions are usually attractive and the population converges around a single position.

\begin{figure}[tbh]
    \centering
    \includegraphics[width=0.6\linewidth]{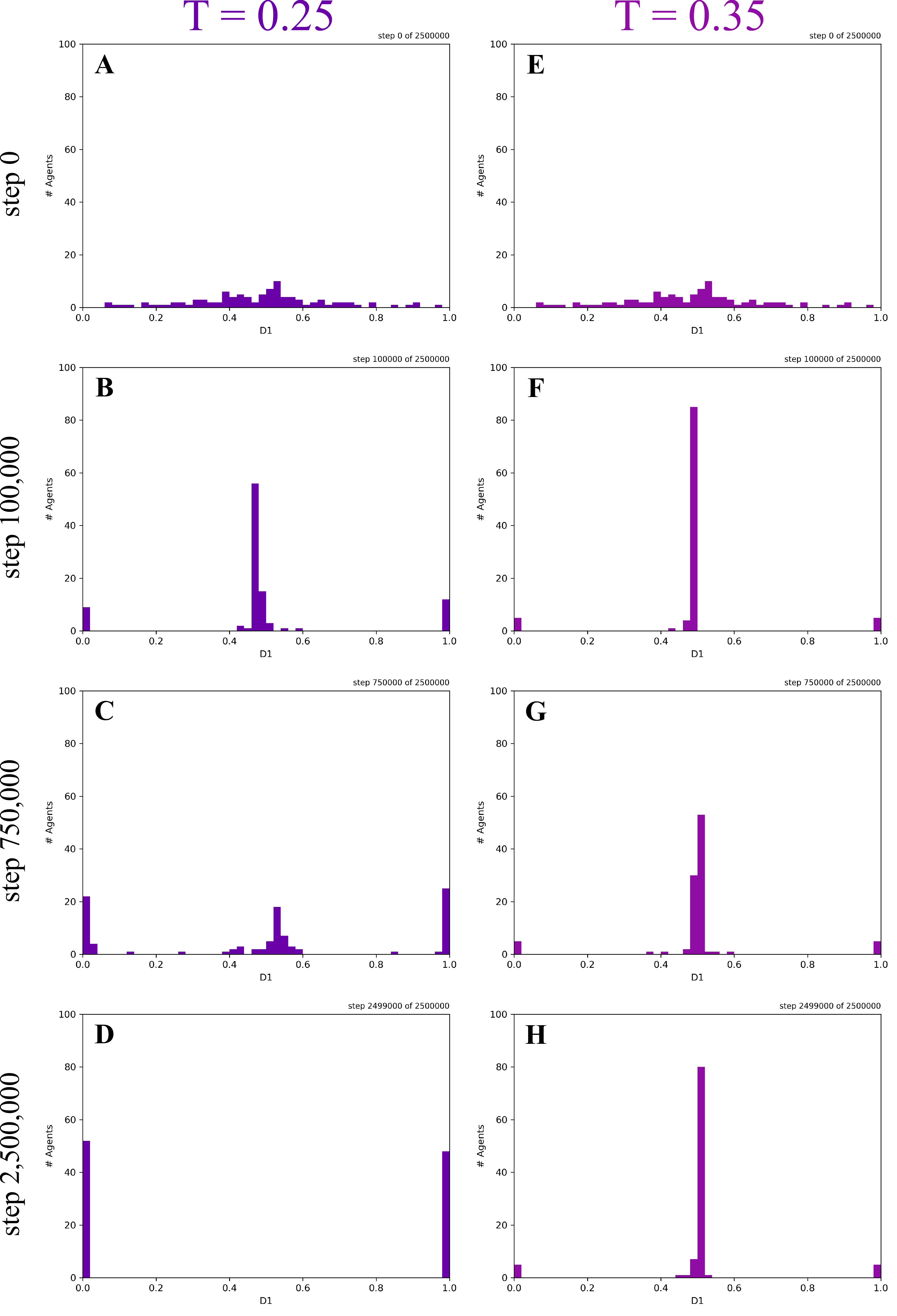}
    \caption{\textbf{With Intermediate Tolerance, Can the Center Hold?}
    Snapshots of the population's ideological positions over time, shown as histograms for the $T = 0.25$ and $T = 0.35$ runs shown in \figtext~\ref{fig:toleranceevo}.
    \textbf{(A)} Initially, the 100 actors are normally distributed with mean 0.5 and standard deviation $\sigma = 0.2$.
    \textbf{(B)} At step 100,000, the $T = 0.25$ run forms a moderate majority of $\sim$80 actors flanked by extreme groups at the far-left and far-right of $\sim$10 actors each.
    \textbf{(C)} The extreme groups grow steadily as the moderate majority dissolves.
    \textbf{(D)} After 2,500,000 steps or --- using our estimation of one interaction per actor per day --- about 70 years, all actors have converged to the extremes in equal proportions.
    \textbf{(E)}--\textbf{(G)} The $T = 0.35$ run forms and maintains a larger moderate majority ($\sim$90 actors) that remains stable over all 2,500,000 steps.
    See Movie S1 for animations.}
    \label{fig:centerhold}
\end{figure}

The interesting cases occur at intermediate levels; two representative examples are $T = 0.25$ and $T = 0.35$.
As \figtext~\ref{fig:toleranceevo} shows, $T = 0.25$ produces polarization that increases slowly at first and then rapidly takes hold and goes to the extreme.
In contrast, $T = 0.35$ produces a nearly constant, low level of polarization that remains stable for the entire simulation.
To understand how the dynamics of these two runs play out so differently, \figtext~\ref{fig:centerhold} shows snapshots of the population's ideological positions at key points in the run.

\figstext~\ref{fig:centerhold}A and~\ref{fig:centerhold}E show the identical initial condition for both runs.
Both the $T = 0.25$ and $T = 0.35$ runs quickly form a majority of moderates near the ideological center flanked by smaller extreme groups at the far-left and far-right (\figstext~\ref{fig:centerhold}B and~\ref{fig:centerhold}F).
These three groups are mutually intolerant and repulse one another.
In the $T = 0.25$ run, repulsion gradually erodes the moderate majority (\figtext~\ref{fig:centerhold}C) until all actors have joined an extremist group (\figtext~\ref{fig:centerhold}D), while in the $T = 0.35$ run the moderate majority remains in dynamic equilibrium among positions $[0.45, 0.55]$ for the entire simulation (\figstext~\ref{fig:centerhold}F--H, Movie S1).

This small change in tolerance yields such divergent long-term outcomes because of a subtle and unanticipated effect that we label \textit{repulsive extremism}.
A left-extremist at position 0.0 repulses any actor to the right of position $T$, pushing it further to the right; likewise, a right-extremist at position 1.0 repulses any actor to the left of position $1 - T$, pushing it further to the left.
Moderates that interact repeatedly with the same extreme group may be repulsed out of the center to the opposite extreme, ultimately dissolving the majority (\figstext~\ref{fig:centerhold}B--D).
However, if the repulsive effect leaves the moderates within the tolerance range of the central majority, their mutual attraction, combined with repulsion from the opposing extremists, will reinforce and shift the majority as it attracts and reabsorbs those that were repulsed.
The emergent effect is one in which the majority reaches and remains in what appears to be a dynamic equilibrium of centrist positions (\figstext~\ref{fig:centerhold}F--H, Movie S1), maintaining a diversity of opinion over long time spans.
Without repulsive extremism, the population would instead converge to a single ideological position from which it would never move.

The determining factor in the long-term effect of repulsive extremism, then, is frequency of interaction between extremists and moderates.
This relies critically on the relative size of the extreme groups compared to the moderates since pairs of actors are chosen uniformly at random to interact.
Tolerance plays a key role in determining these sizes: larger $T$ leads to larger attraction which kickstarts the initial concentration of actors near the center into a strong central majority, while the few remaining actors are repulsed and form small extremist groups.
When the extremist groups are small enough relative to the moderate majority, as in the $T = 0.35$ run, the center holds.
With larger extremist groups such as those in the $T = 0.25$ run, there is higher probability of moderates having repeated interactions with extremists, eventually dissolving the moderate majority.

\boxtext{With low tolerance, even interactions between similar actors can result in repulsion, leaving little hope of avoiding runaway polarization with all actors holding extreme positions.
Sufficiently high tolerance, on the other hand, leads to consensus at a single ideological position.
At intermediate levels, a moderate majority can form that is flanked by repulsive extremists.}

\subsection*{Responsiveness}

Actors can be more or less \textit{responsive} to interactions, depending on how far they move when attracted or repulsed by another.
Recall that the AR rule says that if an actor tolerates another's position, the actor moves a fraction $R$ of the distance between them toward the other, and otherwise moves the same distance away from the other.
Thus, larger values of $R$ represent increased responsiveness.

\begin{figure}[tbh]
    \centering
    \includegraphics[width=0.6\linewidth]{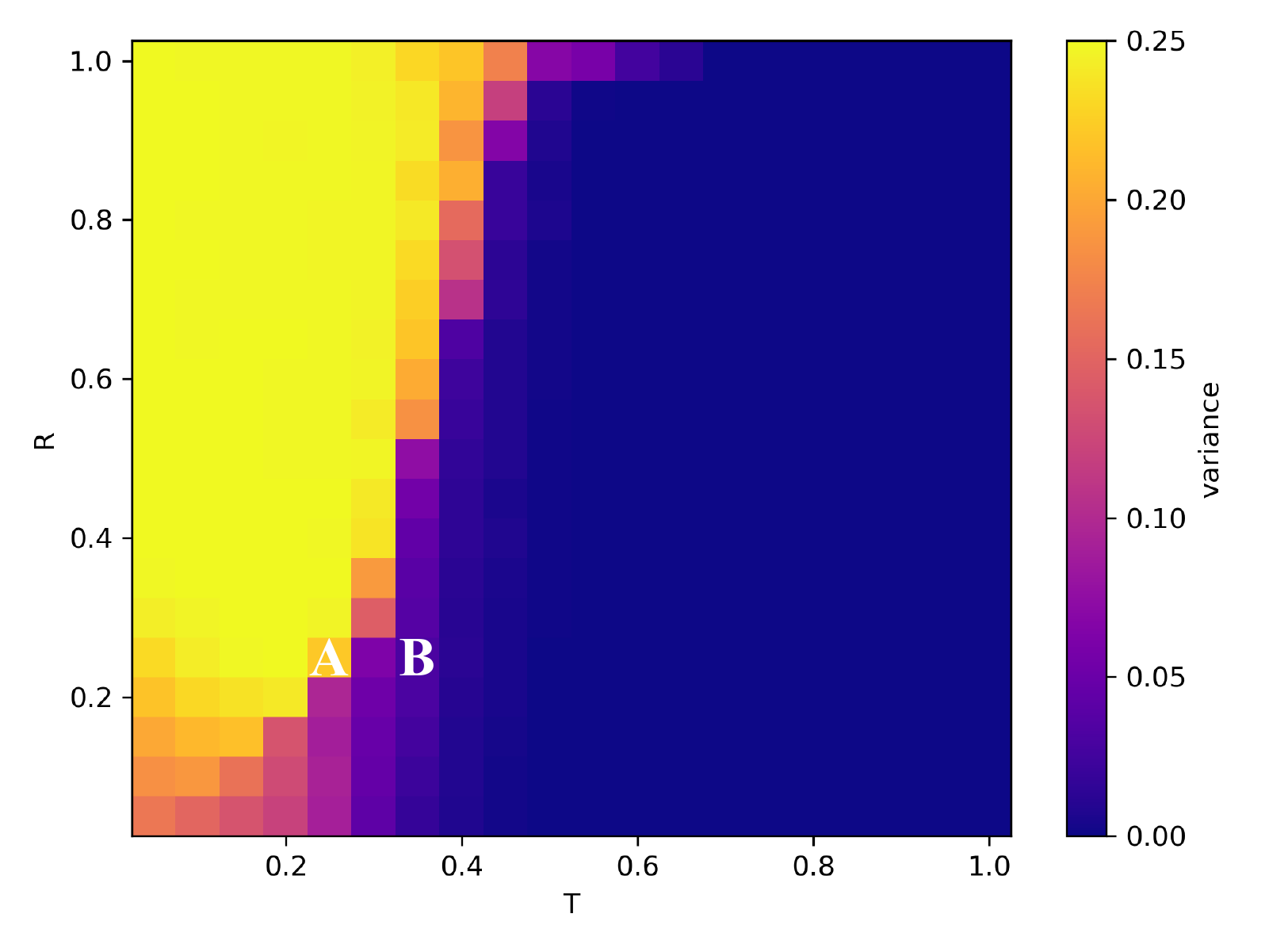}
    \caption{\textbf{The Effects of Responsiveness (R) as a Function of Tolerance (T).}
    Average polarization of the population's ideological positions after 1,000,000 steps, averaged over 20 iterations for each $(T, R)$ pair.
    $T$ and $R$ are both varied over the range $0.05, 0.10, \ldots, 1.0$.
    There is a phase change from extreme polarization (yellow) with low $T$ to convergence (dark blue) with high $T$.
    The phase change is largely independent of $R$.
    \textbf{(A)} and \textbf{(B)} indicate the $T = 0.25$ and $T = 0.35$ cases shown in \figtext~\ref{fig:centerhold} on the boundary of the phase change.}
    \label{fig:responsivesweep}
\end{figure}

\figtext~\ref{fig:responsivesweep} shows that the phase change from complete polarization to convergence to a single ideological position, shown in \figtext~\ref{fig:toleranceevo}, is largely invariant with respect to $R$.
Although actors move much more slowly when $R$ is low (e.g., $R \leq 0.2$), it does not change the outcome.
Intermediate polarization occurs on the boundary of the two regimes ($0.25 \leq T \leq 0.45$), as shown in \figtext~\ref{fig:centerhold}.
Interventions focused on responsiveness are thus unlikely to mitigate polarization because outcomes are largely determined by $T$ except when $R$ is very small.

\subsection*{Exposure}

For a variety of reasons, people tend to be exposed less frequently to opinions that are different from their own than they are to similar opinions.
The strength of this tendency is called the population's \textit{exposure}.
Low exposure means the tendency is strong, while high exposure means that actors listen to distant and similar opinions almost equally.
Empirically, exposure is an important factor in both affective and ideological polarization~\cite{Iyengar2019-affective,Mason2016-crosscuttingcalm}.
In the ARM, the interaction rule captures exposure by stating that the probability of an actor interacting with another is halved as the distance between them is doubled, scaled by the halving distance $E$.
Put another way, population exposure increases with $E$.

\begin{figure}[tbh]
    \centering
    \includegraphics[width=0.6\linewidth]{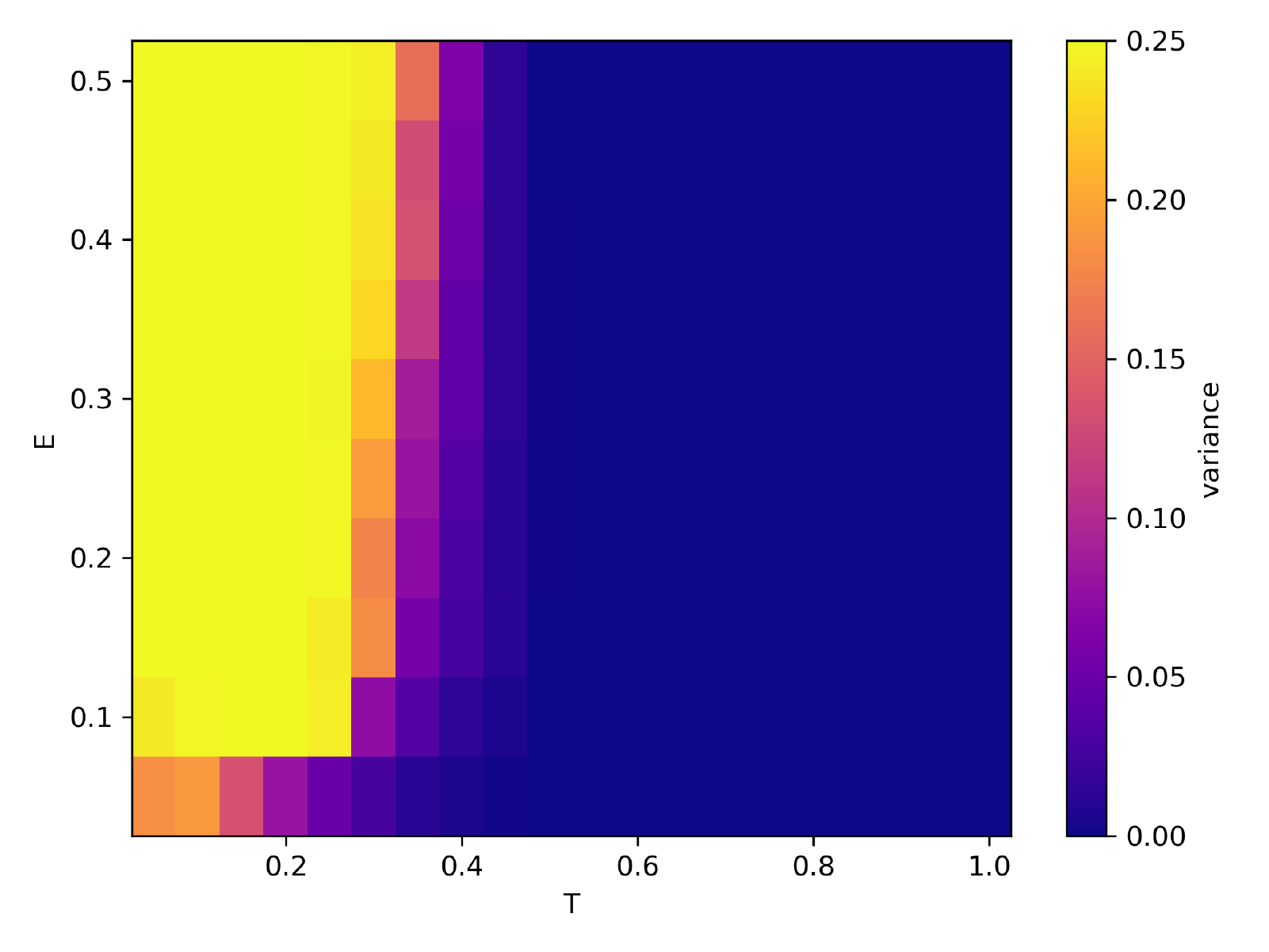}
    \caption{\textbf{The Effects of Exposure (E) as a Function of Tolerance (T).}
    Average polarization of the population's ideological positions after 2,000,000 steps, averaged over 20 iterations for each $(T, E)$ pair.
    Tolerance is varied over $T = 0.05, 0.1, \ldots, 1.0$ and exposure is varied over $E = 0.05, 0.1, \ldots, 0.5$.}
    \label{fig:exposuresweep}
\end{figure}

\figtext~\ref{fig:exposuresweep} shows the effect of different levels of exposure for different levels of tolerance.
For all but the lowest exposures ($E \geq 0.1$), tolerance dominates exposure in determining the population's outcome: just as in \figtext~\ref{fig:responsivesweep}, low tolerance ($T \leq 0.25$) leads to extreme polarization while sufficiently high tolerance ($T \geq 0.5$) yields total convergence.

\begin{figure}[tbh]
    \centering
    \includegraphics[width=0.6\linewidth]{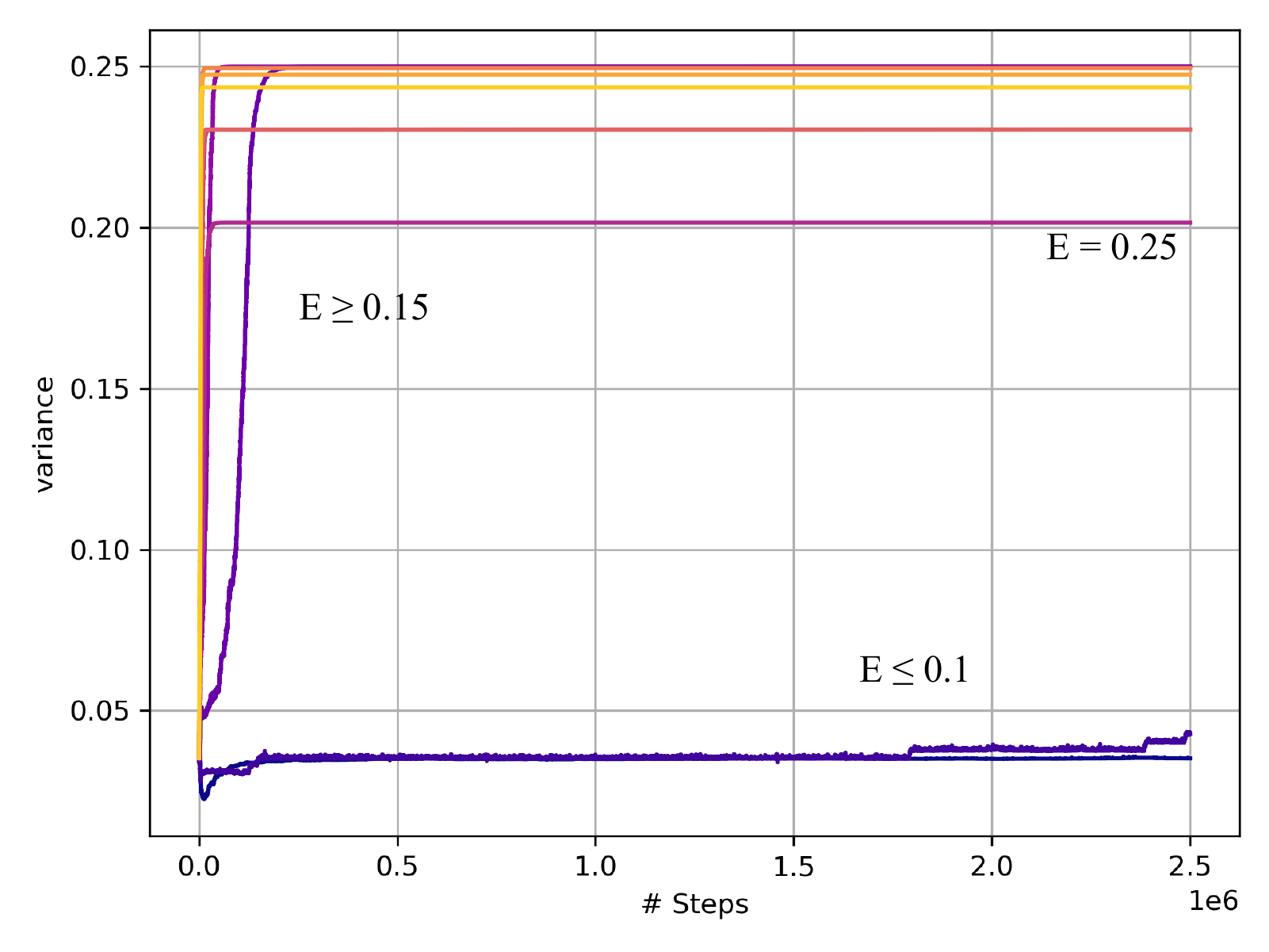}
    \caption{\textbf{The Effects of Exposure (E) for Intermediate Tolerance (T).}
    Polarization of the population's ideological positions over time when $T = 0.3$ is fixed and exposure is varied over the range $E = 0.05, 0.1, \ldots, 0.5$ (dark blue to yellow).
    $E \leq 0.1$ leads to a stable moderate majority flanked by repulsive extremists, while $E \geq 0.15$ leads to rapid polarization.}
    \label{fig:exposureevo}
\end{figure}

At intermediate levels of tolerance ($0.3 \leq T \leq 0.4$), polarization increases with exposure.
To explore this further, we fix $T = 0.3$ and investigate the population's polarization over a longer period for varying degrees of exposure (\figtext~\ref{fig:exposureevo}).
No runs produce convergence, but we observe two distinct types of polarization.
When $E \geq 0.15$, actors often interact with and are repulsed by those who have dissimilar opinions, quickly leading to extreme polarization.
Maximum population variance is not always achieved because the extreme groups may have asymmetric sizes (see, e.g., Movie S2 where the $E = 0.25$ run produces $30\%$ left-extremists and $70\%$ right-extremists for a variance of $\sigma^2 \approx 0.2$).
On the other hand, when actors rarely interact with anyone with different opinions ($E \leq 0.1$), the population can maintain a stable moderate majority flanked by extremist groups for many time steps (Movie S2).

\boxtext{When a population is stubbornly intolerant, the majority of interactions are repulsive.
Increasing exposure only makes these interactions more likely, leading to polarization with rapid adoption of extreme positions.
Low exposure, however, preserves clusters of like-minded individuals by greatly decreasing the probability of repulsive ``cross-cultural'' interactions, inhibiting polarization.}

\subsection*{Multiple Ideological Dimensions}

The previous investigations can be extended to a setting with multiple ideological dimensions.
For simplicity, we consider a two-dimensional ideological space modeled as the unit square in which each actor has an ideological position ranging from 0.0 to 1.0 on each dimension.
The AR rule parameterized by tolerance $T$ and responsiveness $R$ remains the same, but with pairwise distances between actors computed in 2D Euclidean space.
Note that in two dimensions, the maximum distance between two ideological positions is $\sqrt{2} \approx 1.414$ and the maximum variance is $\sigma^2 = 0.5$.

\boxtext{Tolerance and responsiveness have similar effects in two ideological dimensions (\figtext~\ref{fig:responsivesweep2D}) as in one (\figtext~\ref{fig:responsivesweep}), yielding a phase change from extreme polarization with low tolerance to consensus with sufficiently high tolerance.}

With two ideological dimensions, we can investigate populations that have different exposures per dimension.
We generalize the interaction rule to consider two exposures, $E_1$ and $E_2$.
Similar to the one-dimensional case, with high $E_1$ and $E_2$ actors often interact with and are repulsed by others beyond their tolerance, causing extreme polarization (\figtext~\ref{fig:exposuresweep2D}).
However, populations with low exposure even on just one dimension avoid extreme polarization for most degrees of exposure on the other dimension.

\begin{figure}[tbh]
    \centering
    \includegraphics[width=0.8\linewidth]{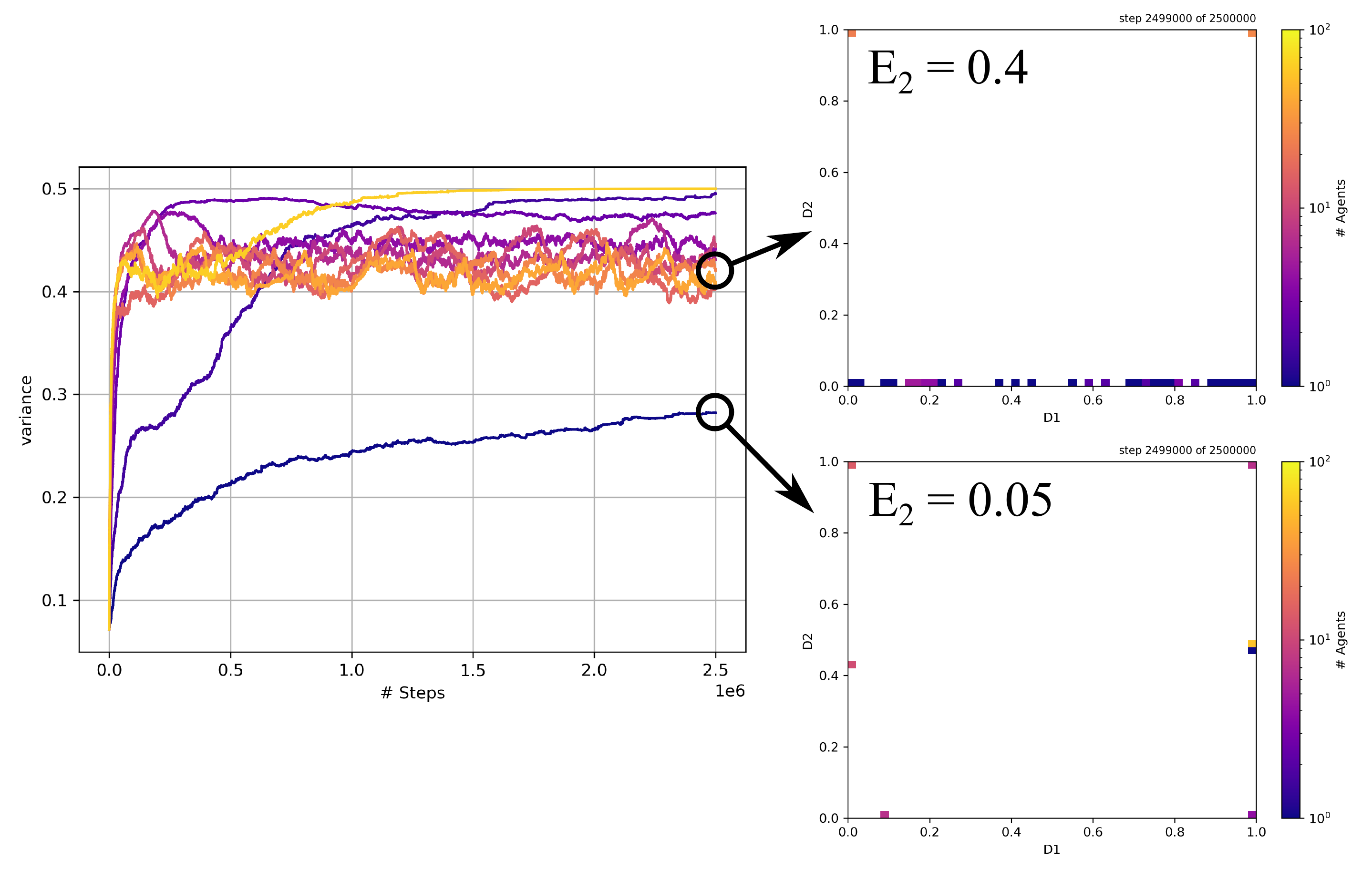}
    \caption{\textbf{Avoiding Maximum Polarization with Low Exposure (E).}
    Polarization of the population's ideological positions over time when exposure on the first ideological dimension is fixed at $E_1 = 0.1$ and exposure on the second ideological dimension is varied over $E_2 = 0.05, 0.1, \ldots, 0.5$ (dark blue to yellow).
    Insets: Final configurations of the population after 2,500,000 steps for the $E_2 = 0.4$ (top) and $E_2 = 0.05$ (bottom) runs as a 2D histogram whose colors indicate concentrations of actors on a log-scale.
    See Movie S3 for animations.}
    \label{fig:exposureevo2D}
\end{figure}

To better understand how low exposure on just one dimension can help mitigate polarization, we consider the situation where actors with disparate views on the first ideological dimension are less likely to interact ($E_1 = 0.1$) while varying $E_2$.
\figtext~\ref{fig:exposureevo2D} shows two distinct population behaviors.
When $E_2 = 0.05$, intermediate polarization occurs slowly.
This effect is a generalization of repulsive extremists to two dimensions: mutually repulsive groups emerge and move to positions at the boundaries and corners of ideological space that are the most stable, interacting only occasionally with other groups (\figtext~\ref{fig:exposureevo2D}, bottom inset).

The second behavior shown in \figtext~\ref{fig:exposureevo2D} is large but oscillating polarization occurring when $0.2 \leq E_2 \leq 0.45$.
As an example, the top inset of \figtext~\ref{fig:exposureevo2D} shows the final configuration of the $E_2 = 0.4$ run.
The population has polarized completely along the second dimension with all actors on either the top or bottom edge due to high exposure on that dimension enabling many repulsive interactions.
Surprisingly, the same is not true of the first dimension.
After 2,500,000 steps, only the top edge has polarized into opposite corners; actors on the bottom edge are distributed bimodally.
The left- and right-leaning bottom actors repulse each other as in one-dimensional settings, but unlike in one dimension, they maintain a diversity of ideological positions.
Whenever a bottom actor gets close to the bottom-right corner, for example, the first dimension's low exposure causes it to interact frequently with the top-right corner of actors.
These two right corners are mutually intolerant and repulsive, causing the bottom actor to be pushed back out of the bottom-right corner.
This subtle and surprising effect shows that frequent interactions between actors close to one another on the low exposure dimension but potentially very far from each other on the high exposure dimension can cause oscillations on the low exposure dimension that remain stable for long time periods.

Although asymmetric exposures yield interesting new behaviors, the key takeaway for higher dimensions is analogous to what was observed in one dimension.
Low exposure can limit interactions between mutually intolerant groups, enabling the population to avoid a flurry of repulsive interactions that can dissolve moderate clusters.
Extreme polarization is avoided as long as these moderate clusters persist, as shown in \figtext~\ref{fig:exposureevo2D}.

\subsection*{Economic Self-Interest}

We next consider a one-dimensional variant of the ARM in which each actor has a preferred ideological position based on \textit{economic self-interest}.\footnote{We describe the preference as due to economic self-interest, but it could represent many other reasons for an actor's innate preference for a certain position.}
For example, an actor with low income might have an economic self-interest in a position to the left of the median.
Although income, wealth and many other attributes are highly skewed, it is not unreasonable to assume that positions follow a normal distribution (see \figtext~\ref{fig:empiricalinit} and the Supplementary Materials for a discussion of empirical ideological data).
Therefore, we assume that an actor's initial position represents its preferred position.
We model the effect of economic self-interest by assuming that with fixed probability $P$, an actor is attracted to its preferred position rather than interacting with another actor.
Thus, $P$ can be thought of as the strength of one's own self-interest in comparison to the effects of other actors in the population.

\begin{figure}[tbh]
    \centering
    \includegraphics[width=0.8\linewidth]{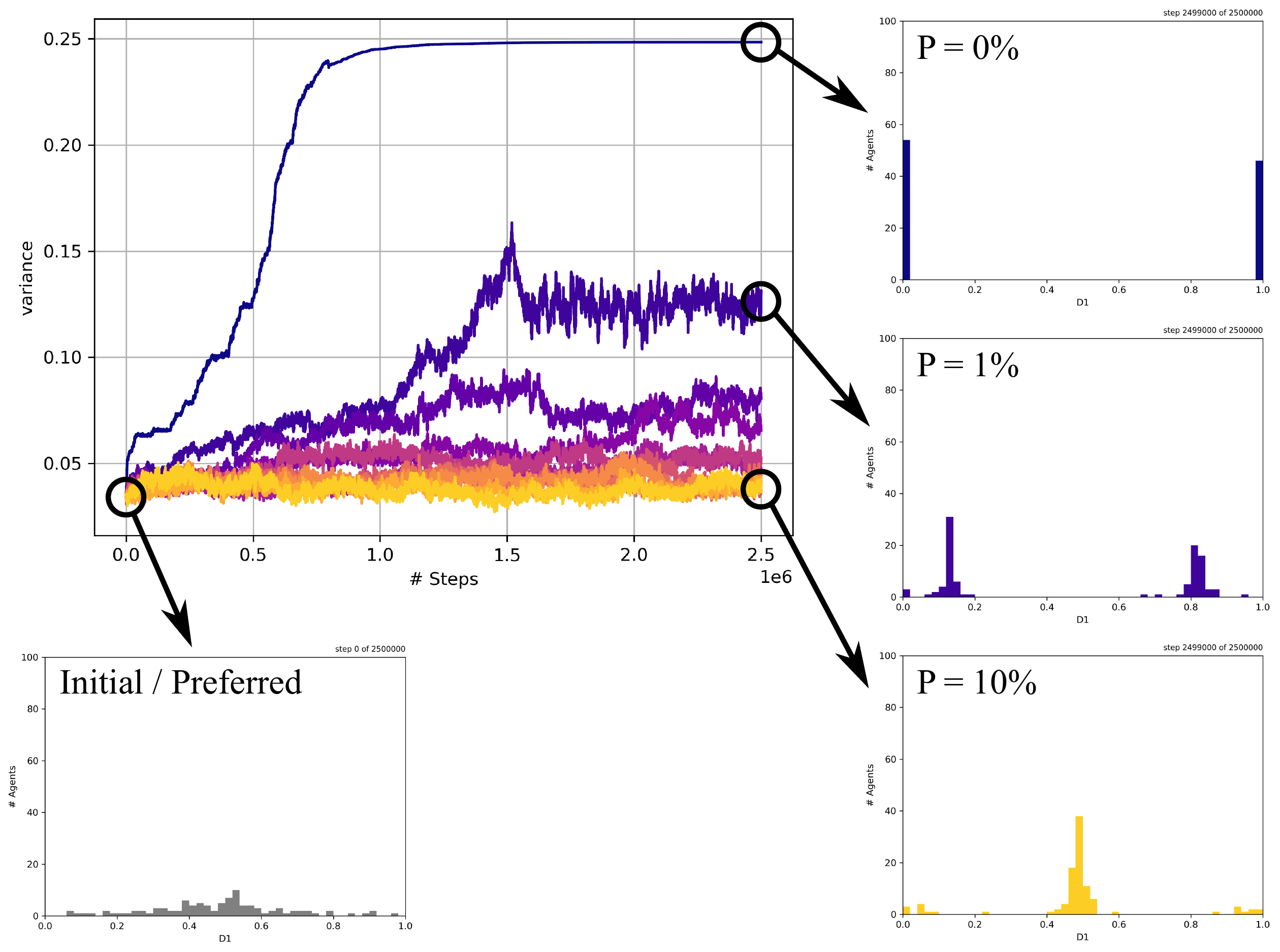}
    \caption{\textbf{The Effects of Economic Self-Interest (P).}
    Polarization of the population's ideological positions over time with varying levels of economic self-interest, $P = 0\%, 1\%, \ldots, 10\%$ (dark blue to yellow) that an actor will be attracted to its preferred (initial) position.
    Left inset: The initial normal distribution of actors' ideological positions, which also represent their preferred positions when acting in self-interest.
    Right insets: Final configurations of the population after 2,500,000 steps for $P = 0\%$, $1\%$, and $10\%$.
    See Movie S4 for animations.}
    \label{fig:selfinterestevo}
\end{figure}

\figtext~\ref{fig:selfinterestevo} shows polarization over time for $P$ ranging from $0\%$ to $10\%$.
The top line is the familiar case when self-interest has no effect ($P = 0\%$) and the population polarizes to the extreme.
But even a very small increase in $P$ avoids extreme polarization: with $P = 1\%$, the population's polarization is roughly halved.
As $P$ increases ($4\% \leq P \leq 10\%$), the population rarely strays from the low level of polarization present in the initial normal distribution.

Without self-interest ($P = 0\%$), the population polarizes in roughly equal proportions to each extreme (\figtext~\ref{fig:selfinterestevo}, top inset).
We use this run as a baseline to understand the relative impact of increasing self-interest.
At very low levels ($P = 1\%$, middle inset), self-interest does not prevent the formation of two mutually repulsive clusters, but it still has a moderating influence.
Because actors are attracted to their more moderate preferred positions, the runs produce bimodal distributions with peaks that oscillate in the $[0.1, 0.2]$ and $[0.8, 0.9]$ ranges, respectively, avoiding extreme polarization.
A ten-fold increase in self-interest ($P = 10\%$, bottom inset) maintains the center and avoids polarization.
The combination of self-interest with attracting interactions produces a tighter concentration of actors near the center than the initial/preferred normal distribution (bottom-left inset).

\boxtext{Even a small amount of self-interest, which biases actors towards their initial positions, can dramatically reduce polarization (\figstext~\ref{fig:selfinterestevo} and~\ref{fig:selfinterestsweep}).}

\subsection*{External Shock}

Our final experiments consider \textit{external shocks} that exogenously shift actors' ideological positions so they become less polarized as they unify around a common problem.
Four examples of such shocks are:
(\textit{i}) \textit{Climate Change Awareness.} As more people experience the reality of increasingly frequent and damaging natural disasters, they may become more willing for the government to take costly action to mitigate its effects.
(\textit{ii}) \textit{The COVID-19 Pandemic.} As infection rates and death counts soar, people may become more willing to support actions such as mask mandates, shutdowns, and emergency funding for testing, contact tracing, and vaccine deployment.
(\textit{iii}) \textit{Economic Recession.} When economic downturns such as the Great Depression, the 2008 recession, and the 2020 economic crisis occur, people become more willing to accept large budget deficits to stimulate the economy.
(\textit{iv}) \textit{War.} When a country declares war or an existing war escalates, people become more willing to pay higher taxes to support the military.

\begin{figure}[tbh]
    \centering
    \includegraphics[width=0.7\linewidth]{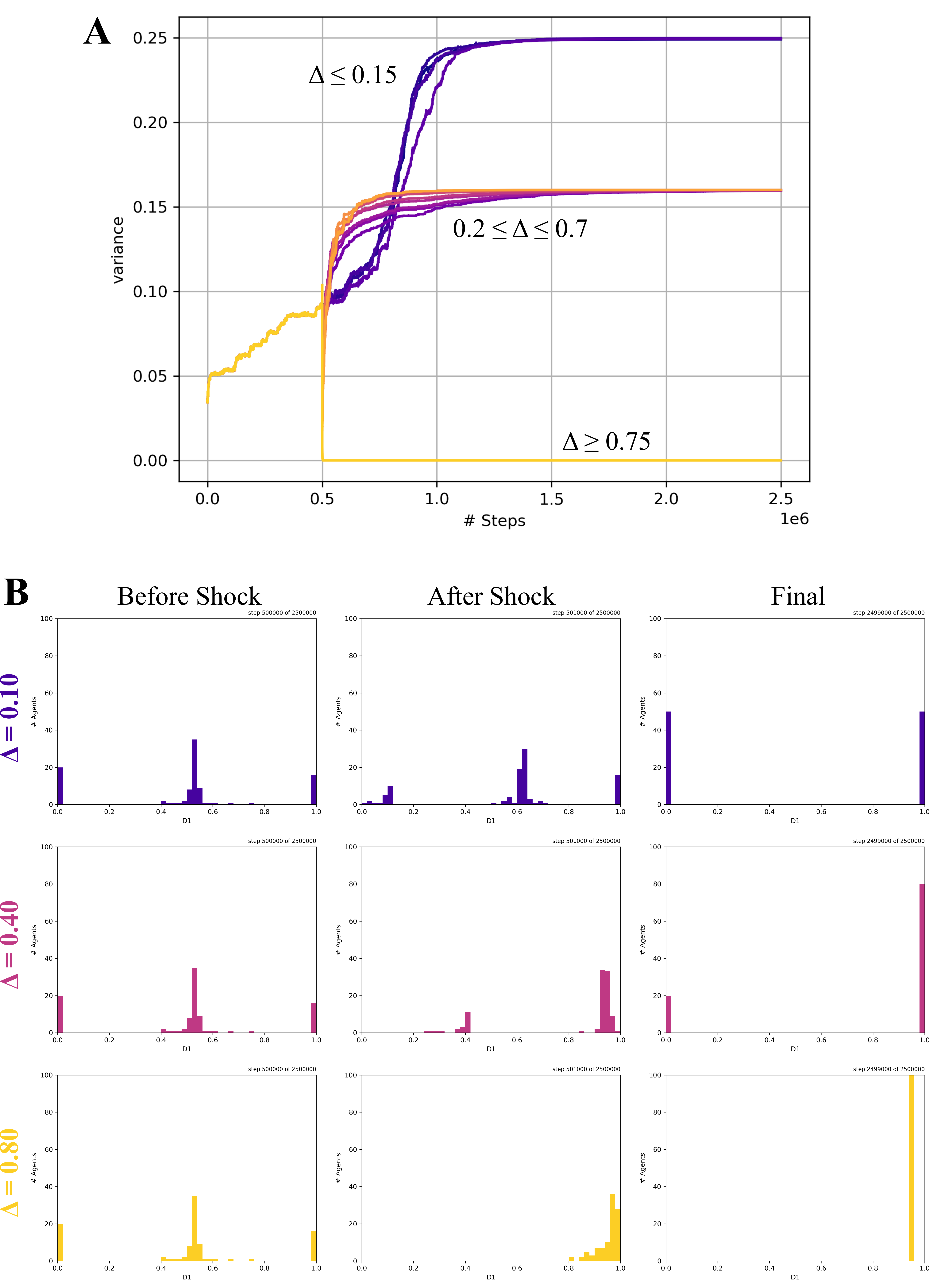}
    \caption{\textbf{The Effects of External Shock Strength ($\Delta$) on Repulsive Extremists.}
    \textbf{(A)} Polarization of the population's ideological positions over time with external shocks of varying strengths $\Delta = 0.0, 0.05, \ldots, 0.8$ (dark blue to yellow) introduced at step 500,000.
    \textbf{(B)} Snapshots of the population's ideological positions as histograms for the $\Delta = 0.1$, $0.4$, and $0.8$ runs just before the shock (step 500,000), shortly after the shock (step 501,000), and in the final configuration (step 2,500,000).
    See Movie S5 for animations.}
    \label{fig:shock}
\end{figure}

In the ARM, an external shock has a strength $\Delta$ and a time (i.e., step) at which it occurs.
At the shock's specified step, all actors' ideological positions are shifted to the right by the shock strength $\Delta$, subject to the usual constraint that no actor can have a position greater than 1.0.
We use the default parameters (Table~\ref{tab:parameters}) to investigate the effects of external shocks of varying strengths introduced at step 500,000.
\figtext~\ref{fig:shock}A shows three distinct outcomes stemming from the same underlying effect: either the shock strength $\Delta$ is large enough to shift ideological groups that were previously repulsive within one another's range of tolerance, enabling them to attract and eventually merge, or the distinct ideological groups remain mutually intolerant and repulse one another.

In more detail, recall that runs with default parameters quickly form a moderate majority flanked by small groups of repulsive extremists (\figtext~\ref{fig:shock}B, left).
Weak shocks ($\Delta \leq 0.15$) shift the left-extremists and moderates to the right by a small amount, but the right-extremists are already at the maximum (\figtext~\ref{fig:shock}B, $\Delta = 0.10$).
So all three groups remain mutually repulsive, the moderate majority is eventually dissolved, and the population converges to extreme polarization as if the shock never happened.
At the other end of the spectrum, very strong shocks ($\Delta \geq 0.75$) make all actors mutually attractive, leading to consensus (e.g., \figtext~\ref{fig:shock}B, $\Delta = 0.8$).
Intermediate shock strengths ($0.2 \leq \Delta \leq 0.7$) weaken intolerance between the moderates and right-extremists but are not strong enough to do so for the moderates and left-extremists (\figtext~\ref{fig:shock}B, $\Delta = 0.40$).
The left-extremists repulse the now-merged moderates and right-extremists, resulting in $\sim$20 actors at the extreme left and the remaining $\sim$80 at the extreme right.\footnote{Since all runs in \figtext~\ref{fig:shock} use the same initial distribution and random seed, all intermediate shock strengths lead to the same proportion of left- and right-extremists.}

\begin{figure}[tbh]
    \centering
    \includegraphics[width=0.6\linewidth]{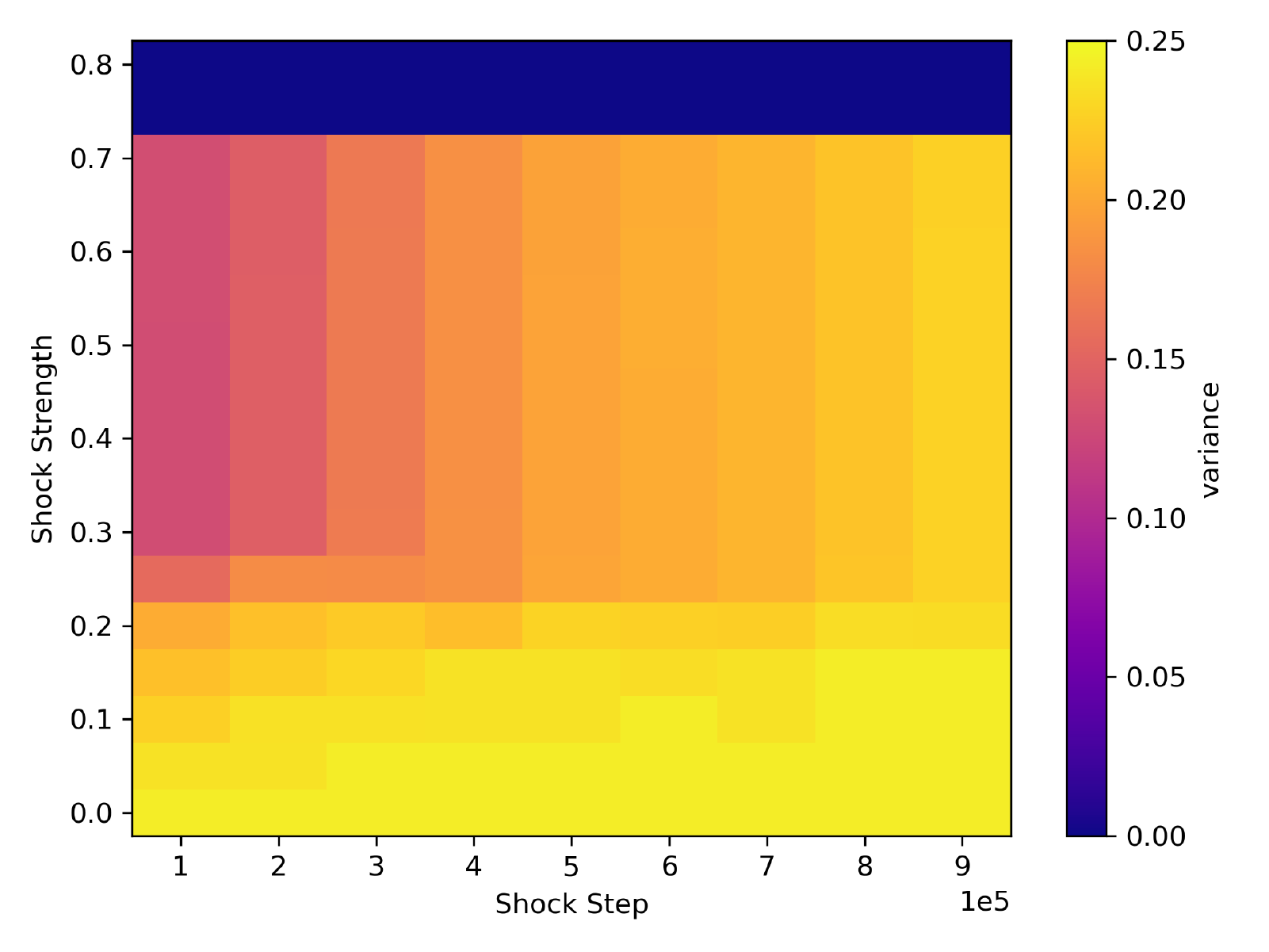}
    \caption{\textbf{The Effects of External Shock by Time and Strength.}
    Average polarization of the population's ideological positions after 2,000,000 steps, averaged over 20 iterations for each $(\Delta, \text{step})$ pair.
    Shocks vary in strength over $\Delta = 0.0, 0.05, \ldots, 0.8$ and are introduced at steps 100,000, 200,000, $\ldots$, 900,000.}
    \label{fig:shocksweep}
\end{figure}

We next consider the effect of introducing shocks at different times.
\figtext~\ref{fig:shocksweep} shows that the three phases of weak, intermediate, and strong shocks observed in \figtext~\ref{fig:shock} occur regardless of when the shock is applied.
Medium strength shocks lead to intermediate polarization, with earlier shocks resulting in smaller variance.
However, no medium strength shock avoids polarization entirely once groups of extremists have already formed.
In particular, if the left-extremists remain repulsive with all other actors after the shock (which are necessarily further to the right), they will be repulsed back to the left extreme.
The timing of this type of shock only affects how large the consolidation of moderate and right-leaning actors is, which affects the final level of polarization.

\boxtext{An external shock, which moves all actors in the same direction, only reduces polarization if it occurs so early that extremist groups have not yet formed or is strong enough to move all actors within a single tolerance distance.}

\section*{Model Extensions}

The ARM can facilitate many natural extensions to study other aspects of polarization, including asymmetry, elites, affective polarization, politics, geography, and other interventions.

\textit{Asymmetry.} To explore some of the asymmetries in current American politics~\cite{Levitsky2018-democraciesdie}, one could use asymmetric initial distributions or vary $T$ across actors.

\textit{Elites.} Elites with large social influence can be represented as fixed ideological positions that others occasionally attract to as in economic self-interest, or as actors who are more likely to be selected by others for interaction.

\textit{Affective Polarization.} To study affective polarization, one could assign each actor as Democrat, Republican, or Independent with initial positions normally distributed around their party's mean, e.g., 0.4, 0.5 and 0.7 respectively.
The degree of affective polarization could be represented as the probability that a member of one party will tolerate a member of the same party and be repulsed by a member of the other party, regardless of distance.

\textit{Politics.} Starting with this model of party affiliations, the primary campaigns of the two parties could be represented as selection among members of one's own party, stopping after a given number of steps.
At that point, the candidate for each party would be assigned the median position of that party's members.
The general election campaign would then treat the two candidates as elites, as described above.
A simple way to represent the general election would have each actor vote for its nearest candidate after a fixed number of steps (the campaign), with the candidate with the most votes declared the winner.

\textit{Geography.} To represent a world in which geography limits interactions, assign each actor to a location on a two-dimensional grid.
Then have an actor select another only from their immediate neighborhood, as in~\cite{Axelrod1997-culturemodel}.

\textit{Other Interventions.} Beyond incentives for self-interest or externally applied shocks, one could consider other interventions such as educational campaigns that nudge the population or interventions that affect actors differentially.

\section*{Discussion}

Despite its simplicity, the Attraction-Repulsion Model sheds light on the original questions.
First is the question of when ideological polarization becomes a runaway process leading to extremism.
Intolerance is the key component of runaway polarization observed in our experiments (\figstext~\ref{fig:toleranceevo}--\ref{fig:responsivesweep}), especially when enhanced by high exposure that enables frequent repulsive interactions between dissimilar individuals (\figtext~\ref{fig:exposuresweep}).
These results also generalize to additional ideological dimensions; for example, runaway polarization occurs with low tolerance when both dimensions have high exposure (\figstext~\ref{fig:empiricalinit} and~\ref{fig:stochasticrepulsion}).

Our second question concerns policy: What interventions can prevent extreme polarization?
Sufficiently high \textit{tolerance} can prevent or dramatically slow polarization (\figstext~\ref{fig:toleranceevo} and~\ref{fig:centerhold}).
For example, when $T = 0.25$, the population polarizes but only very slowly; the center forms but does not hold.
A small increase in tolerance can prevent runaway polarization, but the critical value depends on other parameters such as responsiveness (\figtext~\ref{fig:responsivesweep}) and exposure (\figtext~\ref{fig:exposuresweep}).

Interventions focused on \textit{responsiveness} are unlikely to mitigate polarization because outcomes are largely determined by tolerance effects unless responsiveness is very small (\figtext~\ref{fig:responsivesweep}).
Strictly limiting \textit{exposure to dissimilar views}, however, is an effective mechanism for avoiding rapid polarization (\figstext~\ref{fig:exposuresweep} and~\ref{fig:exposureevo}).
This may at first appear contrary to practical experience: encouraging interactions among those with different views might be expected to decrease polarization by fostering increased tolerance.
However, the ARM treats tolerance and exposure as independent features of a population.
If a population is stubbornly intolerant and on a trajectory to runaway polarization, low exposure decreases the probability of interactions between mutually intolerant groups.
This in turn preserves ideological diversity by inhibiting repulsion.

These results help resolve the controversy about whether contact between different groups tends to increase or decrease their hostility~\cite{Allport1954-natureprejudice,Pettigrew2006-intergroupcontact}.
For example, school desegregation brought people of different races together expecting that exposure to others would reduce hostility.
However, experience in places like Boston in the 1970s~\cite{Formisano2004-bostonagainstbusing} show the opposite because the differences were too great and may have exacerbated preexisting hostility.

When there are two ideological dimensions, extreme polarization can be avoided if the population has low exposure on either issue (\figtext~\ref{fig:exposuresweep2D}).
In other scenarios, however, we observe a ``durable correlation of multiple dimensions.''
This so-called collapse of dimensions can be a threat to democracy, as observed by James Madison in Federalist No.\ 10~\cite{Madison1787-federalist10}.

Attraction to one's \textit{preferred ideological position} is surprisingly effective at preventing polarization, even if this preference is very small compared to the influence of others.
Under these conditions, the population tends to moderate (\figstext~\ref{fig:selfinterestevo} and~\ref{fig:selfinterestsweep}).
This is perhaps the most promising result of the model because it suggests a direction for policy intervention by which a polarizing dynamic could be moderated.

One might imagine that an \textit{external event} that moves everyone in one direction could be another mechanism for controlling polarization, but our model shows that such a shock has to be surprisingly large to succeed (\figtext~\ref{fig:shock}).
If the population is on a polarizing trajectory, the sooner the shock occurs, the more likely it is to have an effect (\figtext~\ref{fig:shocksweep}).
If the population is already somewhat polarized, a weak shock does not stop the process, although a strong shock can.
Even a medium shock (e.g., to the right) only changes the relative sizes of extreme groups, but in the end all actors become extremists, with moderates and right-extremists combined and polarized from the left-extremists (\figtext~\ref{fig:shock}B, middle row).
In future work, it will be interesting to compare these unidirectional shocks to those that produce centrist, ``rally around the flag'' effects.

Without intervention, the ARM typically exhibits either extreme polarization or complete convergence (\figstext~\ref{fig:responsivesweep} and~\ref{fig:exposuresweep}).
Although we have focused on preventing extreme polarization, we note that some diversity of opinions is bound to exist in any open society and is likely necessary for holding elites accountable and sustaining healthy democracy.
In the ARM, we observe some scenarios where a few repulsive extremists help reinforce the center, allowing a moderate majority to remain stable for long periods of time while continuing to react against extremists (\figstext~\ref{fig:centerhold}E--H).
With too many extremists, however, the cluster of moderates ultimately dissolves (\figstext~\ref{fig:centerhold}A--D).
Ideal levels of ideological diversity in a society thus may be sensitive to factors such as interaction patterns and group structures.

\section*{Conclusion}

With just two simple rules, the Attraction-Repulsion Model yields complex dynamics that provide policy-relevant insights into mechanisms for preventing extreme polarization.
While repulsion is often omitted from models of political polarization, we find that in some circumstances repulsion from those of whom we are intolerant can reinforce a moderate majority.

One advantage of an agent-based model such as the ARM is its ability to generate distributions of possible outcomes on different runs of the model, such as rare ``black swan'' events which are difficult to obtain with equation-based models.
For example, while typical runs with $T = 0.35$ yield low polarization (\figstext~\ref{fig:centerhold}E--H and~\ref{fig:responsivesweep}B), rare initial distributions and interaction patterns can lead to polarization four times as large (Movie S6).
While such events are rare by definition, they are important because they can have large and lasting consequences when they do occur.
An interesting question is how to determine whether particular historical events, such as the rise of Hitler, are black swans or expected outcomes of general dynamical processes.
Because democracies require compromise, which is almost impossible when electorates are deeply divided, understanding the forces that promote or inhibit political polarization is crucial for sustaining democracies when they are challenged---whether by black swans or predictable trends.
Models like those in this Special Issue on the dynamics of polarization can help, especially if they are both simple enough to understand where the results come from and subtle enough to provide relevant new insights.

\section*{Acknowledgments}

We thank D.\ Axelrod, V.\ Axelrod, S.\ Croker, D.\ Kinder, C.\ Tausanovitch, C.\ Warshaw, B.\ Edwards, and the participants in this PNAS Special Feature, especially the organizers H.\ Milner and S.\ Levin.
R.A.\ thanks the University of Michigan for research support.
J.J.D.\ is supported in part by the NSF (CCF-1733680), the U.S.\ ARO (MURI W911NF-19-1-0233), and the ASU Biodesign Institute.
S.F.\ is supported in part by the NSF (CCF-1908633), DARPA (FA8750-19C-0003, N6600120C402), AFRL (FA8750-19-1-0501), and the Santa Fe Institute.

\bibliographystyle{unsrt}
\bibliography{refs}

\begin{thebibliography}{10}

\bibitem{Mason2016-crosscuttingcalm}
Lilliana Mason.
\newblock A cross-cutting calm: How social sorting drives affective
  polarization.
\newblock {\em Public Opinion Quarterly}, 80(S1):351--377, 2016.

\bibitem{Patty2019-moderatesextremists}
John~W. Patty.
\newblock Are moderates better representatives than extremists? a theory of
  indirect representation.
\newblock {\em Penn, Elizabeth Maggie}, 113(3):743--761, 2019.

\bibitem{Madison1787-federalist10}
James Madison.
\newblock Federalist {N}o.\ 10, 1787.

\bibitem{Levitsky2018-democraciesdie}
Steven Levitsky and Daniel Ziblatt.
\newblock {\em How Democracies Die}.
\newblock Crown, 2018.

\bibitem{Iyengar2019-affective}
Shanto Iyengar, Yphtach Lelkes, Matthew Levendusky, Neil Malhotra, and Sean~J.
  Westwood.
\newblock The origins and consequences of affective polarization in the
  {U}nited {S}tates.
\newblock {\em Annual Review of Political Science}, 22:129--146, 2019.

\bibitem{Yang2020-partypolarization}
Vicky~Chuqiao Yang, Daniel~M. Abrams, Georgia Kernell, and Adilson~E. Motter.
\newblock Why are {U.S.} parties so polarized? {A} ``satisficing'' dynamical
  model.
\newblock {\em SIAM Review}, 62(3):646--657, 2020.

\bibitem{McCarty2019-polarization}
Nolan McCarty.
\newblock {\em Polarization: {W}hat Everyone Needs to Know}.
\newblock Oxford University Press, 2019.

\bibitem{DellaPosta2020-oilspillmodel}
Daniel DellaPosta.
\newblock Pluralistic collapse: The ``oil spill'' model of mass opinion
  polarization.
\newblock {\em American Sociological Review}, 85(3):507--536, 2020.

\bibitem{Putnam2020-upswingtogether}
Robert~D. Putnam and Shaylyn~Romney Garrett.
\newblock {\em The Upswing: {H}ow America Came Together a Century Ago and How
  We Can Do It Again}.
\newblock Simon \& Schuster, 2020.

\bibitem{Flache2017-modelssocialinfluence}
Andreas Flache, Michael M{\"a}s, Thomas Feliciani, Edmund Chattoe-Brown,
  Guillaume Deffuant, Sylvie Huet, and Jan Lorenz.
\newblock Models of social influence: {T}owards the next frontiers.
\newblock {\em Journal of Artificial Societies and Social Simulation}, 20(4):2,
  2017.

\bibitem{Flache2018-monoculturediversity}
Andreas Flache.
\newblock Between monoculture and cultural polarization: {A}gent-based models
  of the interplay of social influence and cultural diversity.
\newblock {\em Journal of Archaeological Method and Theory}, 25:996--1023,
  2018.

\bibitem{Castellano2009-statphyssocial}
Claudio Castellano, Santo Fortunato, and Vittorio Loreto.
\newblock Statistical physics of social dynamics.
\newblock {\em Reviews of Modern Physics}, 81(2):591--646, 2009.

\bibitem{Redner2019-realvotermodels}
Sidney Redner.
\newblock Reality-inspired voter models: {A} mini-review.
\newblock {\em Comptes Rendus Physique}, 20(4):275--292, 2019.

\bibitem{Axelrod1997-culturemodel}
Robert Axelrod.
\newblock The dissemination of culture: {A} model with local convergence and
  global polarization.
\newblock {\em Journal of Conflict Resolution}, 41(2):203--226, 1997.

\bibitem{Baldassarri2007-dynamicspolarization}
Delia Baldassarri and Peter Bearman.
\newblock Dynamics of political polarization.
\newblock {\em American Sociological Review}, 72(5):784--811, 2007.

\bibitem{VazMartins2010-massmedia}
T.~Vaz~Martins, M.~Pineda, and R.~Toral.
\newblock Mass media and repulsive interactions in continuous-opinion dynamics.
\newblock {\em Europhysics Letters}, 91(4):48003, 2010.

\bibitem{Flache2011-smallworlds}
Andreas Flache and Michael~W. Macy.
\newblock Small worlds and cultural polarization.
\newblock {\em The Journal of Mathematical Sociology}, 35(1--3):146--176, 2011.

\bibitem{Krause2019-repulsiondebate}
Sebastian~M. Krause, Fritz Weyhausen-Brinkmann, and Stefan Bornholdt.
\newblock Repulsion in controversial debate drives public opinion into
  fifty-fifty stalemate.
\newblock {\em Physical Review E}, 100(4):042307, 2019.

\bibitem{Santos2021-socialmatching}
Fernando~P. Santos, Yphtach Lelkes, and Simon~A. Levin.
\newblock Social matching algorithms and dynamics of polarization in online
  social networks.
\newblock Submitted to this special issue of PNAS, 2021.

\bibitem{Szymanski2021-tippingpoints}
Boleslaw~K. Szymanski, Manqing Ma, Daniel~R. Tabin, Jianxi Gao, and Michael~W.
  Macy.
\newblock Polarization and tipping points.
\newblock Submitted to this special issue of PNAS, 2021.

\bibitem{Liu2015-pullingcloser}
Christopher~C. Liu and Sameer~B. Srivastava.
\newblock Pulling closer and moving apart: {I}nteraction, identity, and
  influence in the {U.S. S}enate, 1973 to 2009.
\newblock {\em American Sociological Review}, 80(1):192--217, 2015.

\bibitem{Bail2018-socialmediapolarization}
Christopher~A. Bail, Lisa~P. Argyle, Taylor~W. Brown, John~P. Bumpus, Haohan
  Chen, M.~B.~Fallin Hunzaker, Jaemin Lee, Marcus Mann, Friedolin Merhout, and
  Alexander Volfovsky.
\newblock Exposure to opposing views on social media can increase political
  polarization.
\newblock {\em Proceedings of the National Academy of Sciences},
  115(37):9216--9221, 2018.

\bibitem{Mas2013-diffwithoutdistance}
Michael M{\"a}s and Andreas Flache.
\newblock Differentiation without distancing. {E}xplaining bi-polarization of
  opinions without negative influence.
\newblock {\em PLoS ONE}, 8(11):e74516, 2013.

\bibitem{Takacs2016-discrepancydisliking}
K{\'a}roly Tak{\'a}cs, Andreas Flache, and Michael M{\"a}s.
\newblock Discrepancy and disliking do not induce negative opinion shifts.
\newblock {\em PLoS ONE}, 11(6):e0157948, 2016.

\bibitem{Schaffner2021-cces2020}
Brian Schaffner, Stephen Ansolabehere, and Sam Luks.
\newblock {Cooperative Election Study Common Content, 2020}.
\newblock Available online at \url{https://doi.org/10.7910/DVN/E9N6PH}, 2021.

\bibitem{Tausanovitch2013-measuringpolicy}
Chris Tausanovitch and Christopher Warshaw.
\newblock Measuring constituent policy preferences in {C}ongress, state
  legislatures, and cities.
\newblock {\em The Journal of Politics}, 75(2):330--342, 2013.

\bibitem{Allport1954-natureprejudice}
Gordon~W. Allport.
\newblock {\em The Nature of Prejudice}.
\newblock Addison-Wesley, 1954.

\bibitem{Pettigrew2006-intergroupcontact}
Thomas~F. Pettigrew and Linda~R. Tropp.
\newblock A meta-analytic test of intergroup contact theory.
\newblock {\em Journal of Personality and Social Psychology}, 90(5):751--783,
  2006.

\bibitem{Formisano2004-bostonagainstbusing}
Ronald~P. Formisano.
\newblock {\em Boston Against Busing: {R}ace, Class, and Ethnicity in the 1960s
  and 1970s}.
\newblock University of North Carolina Press, 2004.

\end{thebibliography}

\clearpage

\renewcommand{\thefigure}{S\arabic{figure}}
\setcounter{figure}{0}

\section*{Supplementary Materials}

Movie S1. With Intermediate Tolerance, Can the Center Hold?

Movie S2. The Effects of Exposure in One Ideological Dimension.

Movie S3. The Effects of Exposure in Two Ideological Dimensions.

Movie S4. The Effects of Economic Self-Interest.

Movie S5. The Effects of External Shock.

Movie S6. Black Swan Runs Producing Rare Behavior.

\subsection*{Formal Specification of the Attraction-Repulsion Model (ARM) and its Simulations}

The following is a formal definition of the ARM and the details of its corresponding simulation implementation.
The Python source code and execution instructions can be found at \url{https://github.com/jdaymude/AttractionRepulsionModel}.

The ARM considers a population of $N \in \mathbb{Z}_+$ actors each of which has an ideological position $\vec{x} = (x_1, \ldots, x_D) \in [0, 1]^D$ where $D \in \mathbb{Z}_+$ is the number of ideological dimensions.
Throughout the remainder of the model specification, we use $||\vec{x}||_2 = \sqrt{\sum_{i=1}^D x_i^2}$ to denote the Euclidean norm of $\vec{x}$.

\paragraph*{Initialization.}

In one dimension ($D = 1$), actors' ideological positions are initialized according to a normal (Gaussian) distributed with mean $0.5$ and standard deviation $\sigma = 0.2$.
For multiple dimensions ($D > 1$), actors' initial ideological positions are initialized according to a multivariate normal distribution with means $(0.5)^D$ and covariance matrix $C$ defined as:
\[C(i, j) = \left\{ \begin{array}{cl}
    0.04 & \text{if $i = j$}; \\
    0 & \text{otherwise.}
\end{array} \right.\]
where $i, j \in \{1, \ldots, D\}$.
This covariance matrix implies that the actors' positions are independent and normally distributed with standard deviation $\sigma = 0.2$ along each dimension.
To ensure the randomly sampled positions are within the boundaries of the ideological space $[0, 1]^D$, we use rejection sampling.
It is worth nothing that truncated normal distributions in one dimension have mathematically defined PDFs and CDFs that lend themselves to direct implementation without the need for rejection sampling, but since those in higher dimensions do not, we implement both cases with rejection sampling for consistency.

\paragraph*{Interaction Rule.}

At each step, a pair of distinct actors $(x, y)$ are chosen uniformly at random to interact, where $x$ is the active actor and $y$ is the passive actor.
Let $\vec{x} = (x_1, \ldots, x_D) \in [0, 1]^D$ denote the position of actor $x$, and likewise $\vec{y}$ for actor $y$.
Each ideological dimension $i$ has an \textit{exposure} $E_i \in \mathbb{R}_+$.
actors $x$ and $y$ interact with probability $(1/2)^{\delta_{xy}}$, where:
\[\delta_{xy} = \sqrt{\sum_{i=1}^D \frac{(x_i - y_i)^2}{E_i^2}}\]
In the one-dimensional setting, this probability of interaction simplifies to:
\[\left(\frac{1}{2}\right)^{|x_1 - y_1| / E_1}\]

\paragraph*{Attraction-Repulsion (AR) Rule.}

The AR rule is parameterized by \textit{tolerance} $T \in [0, \sqrt{D}]$ and \textit{responsiveness} $R \in (0, 1]$.
If two actors $x$ and $y$ interact according to the interaction rule, then the active actor $x$ moves according to the following rules:
\[\vec{x} \gets \left\{ \begin{array}{cl}
    \vec{x} + R(\vec{y} - \vec{x}) & \text{if $||\vec{x} - \vec{y}||_2 \leq T$ (attraction)}; \\
    \vec{x} - R(\vec{y} - \vec{x}) & \text{if $||\vec{x} - \vec{y}||_2 > T$ (repulsion)}.
\end{array} \right.\]
To ensure that repulsion does not move an actor outside the boundaries of the ideological space $[0, 1]^D$, we additionally update each $x_i \in \vec{x}$ as:
\[x_i \gets \max(0, \min(1, x_i))\]

\paragraph*{Economic Self-Interest.}

The self-interest intervention considers a probability $P \in [0, 1]$ of attracting to one's own preferred (initial) position instead of interacting with another actor.
Let $x$ be the active actor (chosen uniformly at random) and let $\vec{x}_0$ denote its initial and preferred position.
With probability $P$, $x$ attracts to $\vec{x}_0$ using:
\[\vec{x} \gets \vec{x} + R(\vec{x}_0 - \vec{x})\]
The Interaction and Attraction-Repulsion rules are then skipped for this step.
With the remaining probability $1 - P$, the self-interest intervention has no effect at this step and $x$ proceeds according to the Interaction and Attraction-Repulsion rules.

\paragraph*{External Shock.}

The external shock intervention considers a shock that occurs at time step $S \in \mathbb{Z}_+$ and has a strength $\vec{\Delta} \in [0, 1]^D$.
At a high level, the shock is implemented such that at step $S$, the position $\vec{x}$ of each actor $x$ is shifted by $\vec{\Delta}$, clipped at the boundaries of the ideological space:
\[\vec{x} \gets \min(1, \vec{x} + \vec{\Delta})\]
The simulation framework only allows one actor to update its position at each step, however, so instead we apply the shock to the $i$-th actor at step $S + (i - 1)$, spreading the shock out over $N$ consecutive steps.
The Interaction and Attraction-Repulsion rules are skipped for these steps.

\paragraph*{Measuring Polarization.}

We measure polarization as the \textit{population variance} of the actors' ideological positions.
In the one-dimensional setting ($D = 1$), the population variance of actors' positions $\{x_1, \ldots, x_N\}$ where each $x_i \in [0, 1]$ is:
\[\sigma^2 = \frac{\sum_{i=1}^N x_i^2}{N} - \left(\frac{\sum_{i=1}^N x_i}{N}\right)^2\]
For multiple dimensions ($D > 1$), there are several possible generalizations of population variance.
We use the trace of the covariance matrix, which is computed as the sum of the variances of each dimension.
This is a suitable metric because each ideological dimension is independent; if they weren't, the determinant of the covariance matrix would be more appropriate.

\paragraph*{Random Number Generation.}

All random number generation is seeded for reproducibility.
When taking average statistics over repeated iterations of the same run (i.e., parameter setting), our framework first uses the seed provided by the user to generate a list of random seeds---one per repeated iteration---that are used consistently across runs.
This ensures that the various outcomes observed between runs are due only to the parameter changes and not to the randomness.

\subsection*{Empirical Initial Distribution}

\begin{figure}
    \centering
    \includegraphics[width=0.5\textwidth]{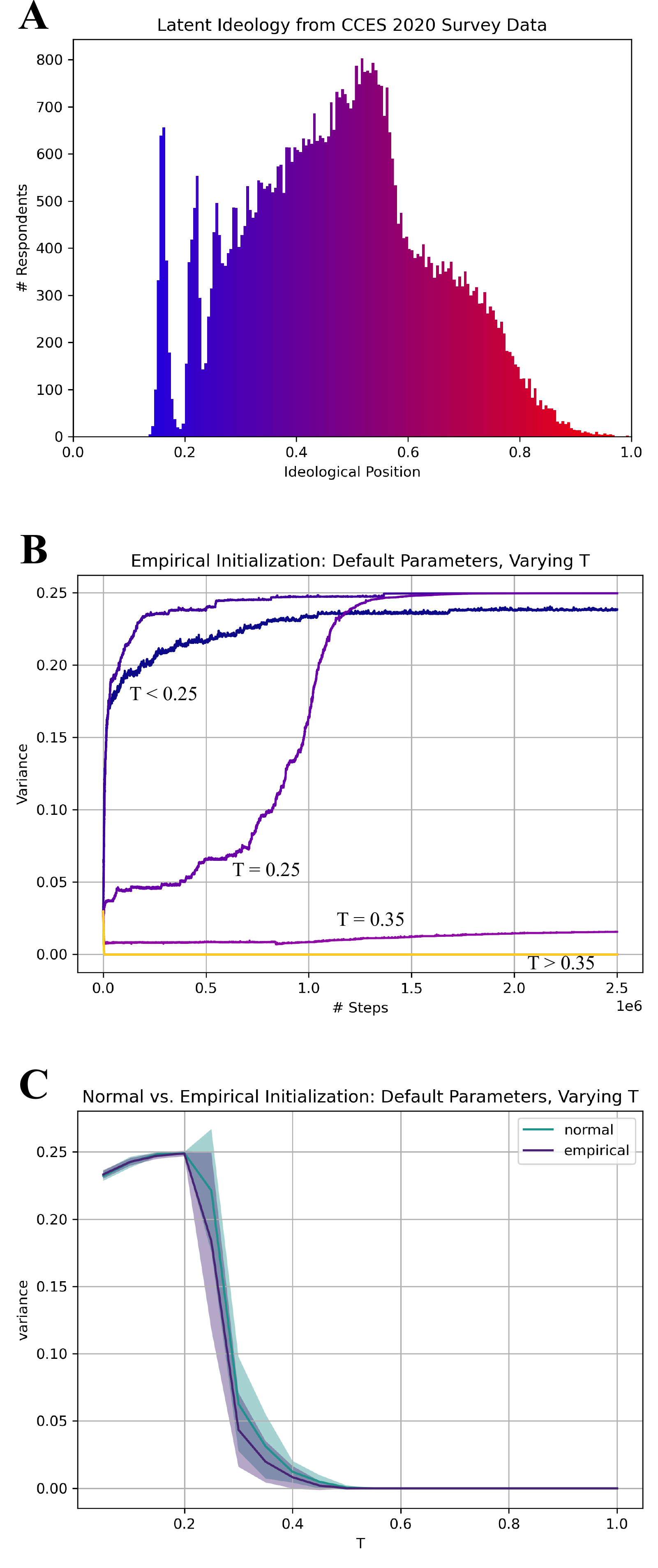}
    \caption{\textbf{The Effects of Empirical Initial Distributions.}
    \textbf{(A)} Estimates of respondents' one-dimensional latent ideology, extracted from the 2020 Cooperative Election Study Common Content.
    \textbf{(B)} Polarization of the population's ideological positions over time when varying tolerance over the range $T = 0.05, 0.15, \ldots, 0.95$ (dark blue to yellow) using empirical initialization.
    \textbf{(C)} Average and standard deviations of polarization of the population's ideological positions after 1,000,000 steps for 20 iterations per tolerance value in $T = 0.05, 0.10, \ldots, 1$ for normally distributed (teal) and empirically distributed (purple) populations.}
    \label{fig:empiricalinit}
\end{figure}

Recall that the populations are initialized according to a normal (Gaussian) distribution with mean $0.5$ and standard deviation $\sigma = 0.2$.
We test the validity of this assumption by comparing the normally-distributed population's behavior to that of a population that is initialized according to an empirical distribution.
The empirical distribution is based on data from the 2020 Cooperative Election Study Common Content~\cite{Schaffner2021-cces2020}, from which respondents' one-dimensional latent ideology can be estimated~\cite{Tausanovitch2013-measuringpolicy}.
The resulting histogram of ideological positions is shown in \figtext~\ref{fig:empiricalinit}A.
Although this empirical distribution has features distinct from the normal (Gaussian) distribution, both are essentially unimodal with a mean near $0.5$.

To generate initial distributions based on this empirical data, actors' initial ideological positions are sampled in $[0, 1]$ according to the weights defined by the empirical histogram.
To avoid any artifacts introduced by sampling from a discrete set of positions (i.e., the histogram's bin midpoints) we apply small, uniformly-sampled noise to each sampled position, approximating a continuous distribution.
Comparing \figtext~\ref{fig:empiricalinit}B to \figtext~\ref{fig:toleranceevo} reveals very little difference between the empirically and normally distributed populations with respect to their polarization over time.
\figtext~\ref{fig:empiricalinit}C shows strong agreement between the average and standard deviations of the empirically and normally distributed populations' polarization levels after 1,000,000 steps over 20 iterations per value of tolerance $T$.
Notably, the empirically distributed populations achieve slightly lower levels of polarization at intermediate tolerance values.
These results validate our choice of normally distributed populations.

\subsection*{Stochastic Attraction-Repulsion}

\begin{figure}
    \centering
    \includegraphics[width=0.46\textwidth]{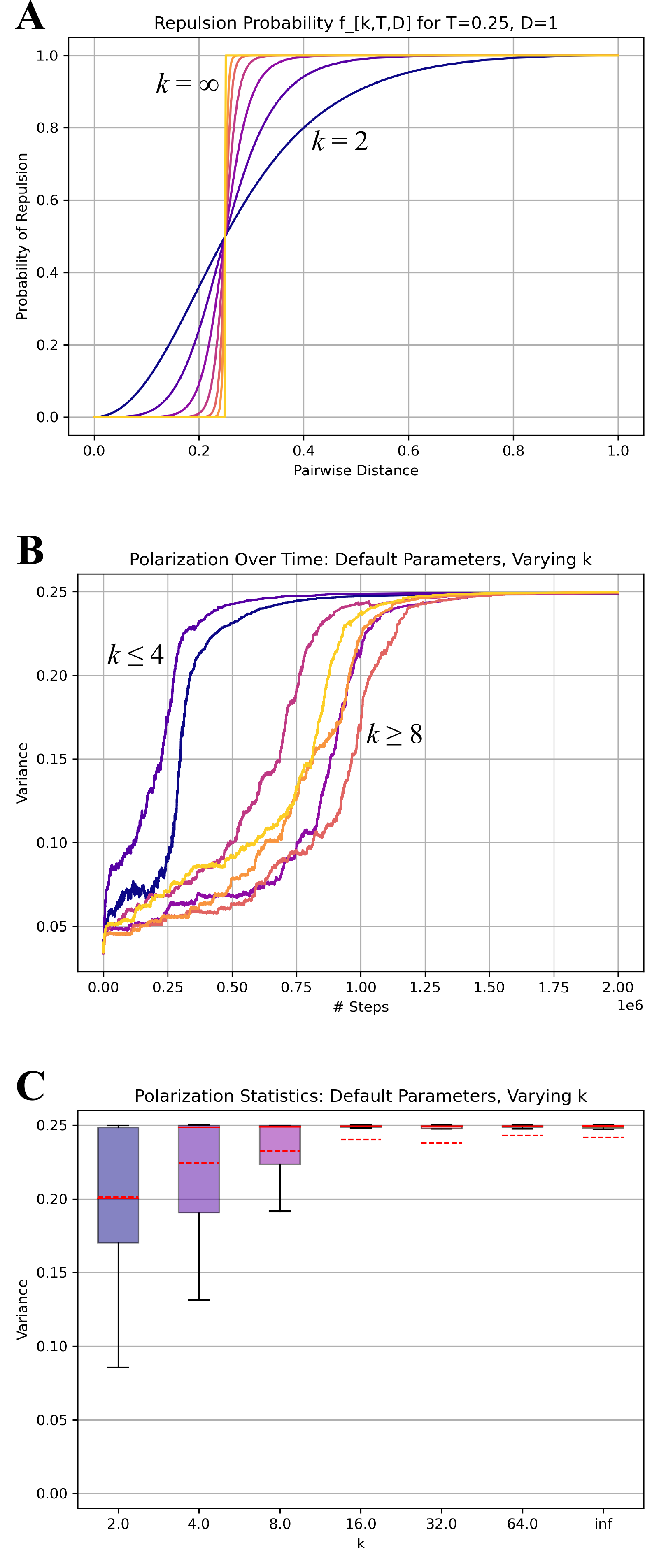}
    \caption{\textbf{The Effects of the Stochastic-Attraction-Repulsion (SAR) Rule of Opinion Change.}
    The SAR rule defines a probability $f_{k,T,D}(d)$ of two interacting actors repulsing when the pairwise distance between their ideological positions is $d$.
    The parameter $k > 1$ controls the steepness, and is varied over the range $k = 2^1, 2^2, \ldots, 2^6, \infty$ (dark blue to yellow) in all three plots.
    \textbf{(A)} The probability of repulsion $f_{k,T,D}(d)$ for varying steepness values $k$, tolerance $T = 0.25$, and number of dimensions $D = 1$.
    \textbf{(B)} Polarization of the population's ideological positions over time for varying steepness values.
    \textbf{(C)} Polarization of the population's ideological positions after 1,500,000 steps for 20 iterations per steepness value.
    Statistics are shown with a standard box plot: each box extends from the lower to upper quartile, whiskers follow Tukey's original definition by extending from the lowest datum above the lower quartile minus $1.5\times$ the interquartile range to the highest datum below the upper quartile plus $1.5\times$ the interquartile range, medians are shown with solid red lines, and means are shown with dashed red lines.
    Some means fall below the whisker range because of outliers that are not shown.}
    \label{fig:stochasticrepulsion}
\end{figure}

Recall that the Attraction-Repulsion (AR) rule causes interacting actors to attract when their pairwise distance is at most the tolerance value $T$ and causes repulsion otherwise.
To validate the robustness of this rule, we compare the resulting behavior to that of an alternative \textit{Stochastic-Attraction-Repulsion (SAR) rule} defined as follows.
If two actors $x$ and $y$ interact according to the interaction rule, then active actor $x$ attracts towards $y$ with probability $1 - f_{k,T,D}(||\vec{x} - \vec{y}||_2)$ and repulses otherwise, where $f_{k,T,D}$ is defined as:
\[f_{k,T,D}(d) = \left(1 + \left(\frac{\sqrt{D}/d - 1}{\sqrt{D}/T - 1}\right)^k\right)^{-1}\]
The function $f_{k,T,D} : (0, \sqrt{D}] \to [0, 1]$ is a logistic function mapping pairwise distances between actors to probabilities of repulsion, where $k > 1$ controls the steepness of the curve, $T \in (0, \sqrt{D})$ is the tolerance parameter, and $D \in \mathbb{Z}_+$ is the number of ideological dimensions (see \figtext~\ref{fig:stochasticrepulsion}A).
This function has several key properties: (\textit{i}) $\lim_{d \to 0}f_{k,T,D}(d) = 0$, i.e., actors occupying the same ideological position should never repulse; (\textit{ii}) $f_{k,T,D}(T) = 1/2$, i.e., actors that are exactly $T$ apart should have a 50\%-50\% chance of attracting or repulsing; and (\textit{iii}) $f_{k,T,D}(\sqrt{D}) = 1$, i.e., actors that are maximally distant should always repulse.
Notably, the AR rule is a special case of the SAR rule obtained by setting $k = \infty$ (\figtext~\ref{fig:stochasticrepulsion}A, yellow).

\figtext~\ref{fig:stochasticrepulsion}B shows a population's polarization over time under the default parameters (Table 1) when using the SAR rule for a range of steepness values $k$.
All runs yield extreme polarization, though those with small steepness values do so more rapidly, suggesting that increased probability of repulsion between similar actors can hasten polarization despite the probability of repulsion between dissimilar actors being decreased.
For larger steepness values, the AR and SAR rules yield similar behavior.
\figtext~\ref{fig:stochasticrepulsion}C further demonstrates that the long-run outcomes (after 1,500,000 steps) for the AR and SAR rules are similar, though the smaller steepness values encode greater stochasticity, allowing for a wider range of intermediate to extreme polarization.

\subsection*{Supplementary Results}

\paragraph*{Model Sensitivity to Core Parameters.}

\begin{figure}
    \centering
    \includegraphics[width=0.5\textwidth]{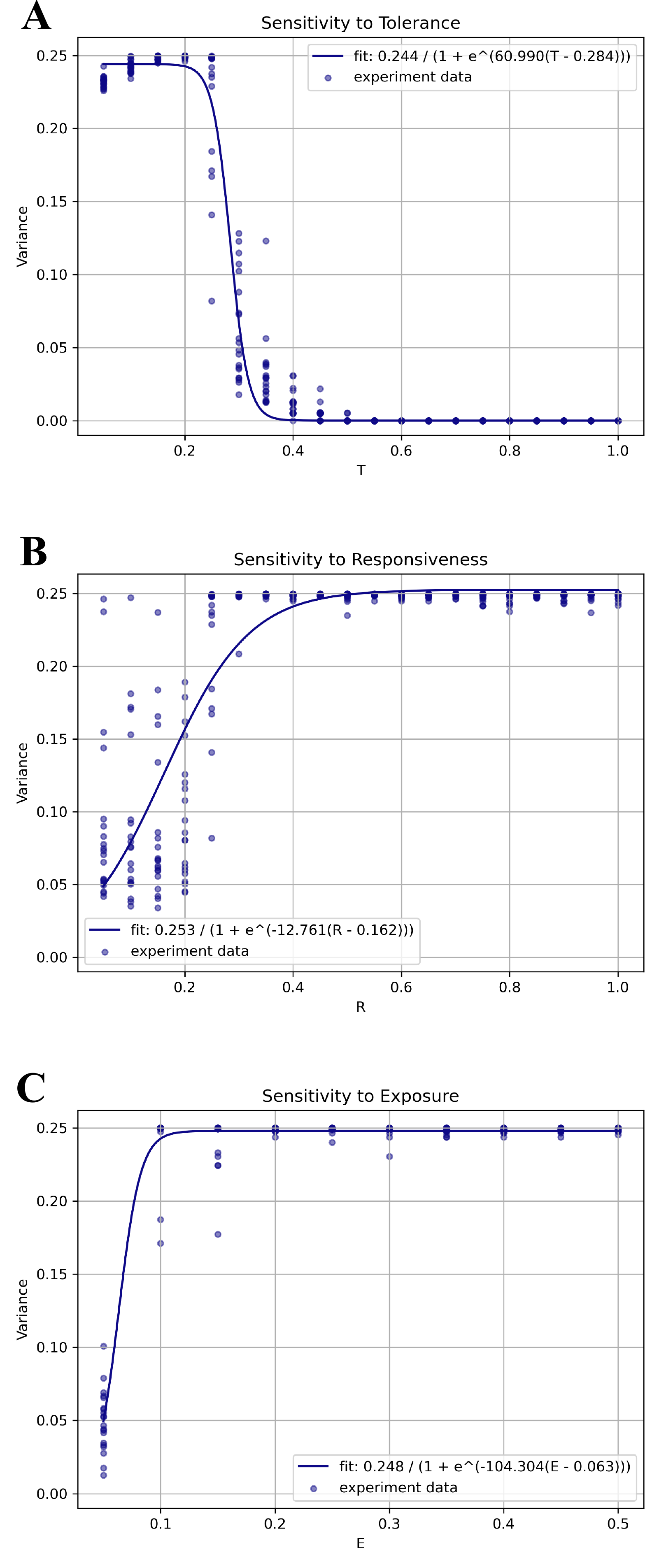}
    \caption{\textbf{Model Sensitivity to Core Parameters.}
    The final polarization of the population's ideological positions for 20 iterations per parameter setting, fit to a logistic curve $f(x) = \frac{a}{1 + e^{-k(x - x_0)}}$.
    \textbf{(A)} For tolerance varied over the range $T = 0.05, 0.10, \ldots, 1.0$, the transition occurs at $T = 0.284 \pm 0.001$ with growth rate $k = -60.990 \pm 3.484$.
    \textbf{(B)} For responsiveness varied over the range $R = 0.05, 0.10, \ldots, 1.0$, the transition occurs at $R = 0.162 \pm 0.005$ with growth rate $k = 12.761 \pm 0.841$.
    \textbf{(C)} For exposure varied over the range $E = 0.05, 0.10, \ldots, 0.5$, the transition occurs at $E = 0.063 \pm 0.001$ with growth rate $k = 104.304 \pm 10.390$.}
    \label{fig:sensitivity}
\end{figure}

We complement \figstext~\ref{fig:responsivesweep} and~\ref{fig:exposuresweep}---i.e., the sweeps of tolerance-responsiveness and tolerance-exposure parameter space---by analyzing the sensitivity of the population's final variance (i.e., polarization) to changes in tolerance $T$, responsiveness $R$, and exposure $E$.
Using the default parameters (Table~\ref{tab:parameters}) and varying one parameter at a time, we find transitions between non-polarized and polarized outcomes (\figtext~\ref{fig:sensitivity}).
To characterize these transitions, we fit the data to a logistic curve $f(x) = \frac{a}{1 + e^{-k(x - x_0)}}$ where $a > 0$ is its maximum value, $k$ is its logistic growth rate, and $x_0$ is its midpoint.
In our context, $k < 0$ indicates a transition from polarized to non-polarized outcomes as the parameter of interest increases, $k > 0$ indicates the opposite (i.e., a non-polarized to polarized transition), and the sensitivity of the transition increases with $|k|$.
As tolerance increases, outcomes transition from polarized to non-polarized at $T = 0.284 \pm 0.001$ with a growth rate of $k = -60.990 \pm 3.484$ (\figtext~\ref{fig:sensitivity}A).
As responsiveness increases, outcomes transition from non-polarized to polarized at $R = 0.162 \pm 0.005$ with a growth rate of $k = 12.761 \pm 0.841$ (\figtext~\ref{fig:sensitivity}B).
Finally, as exposure increases, outcomes transition from non-polarized to polarized at $E = 0.063 \pm 0.001$ with a growth rate of $k = 104.304 \pm 10.390$ (\figtext~\ref{fig:sensitivity}C).
These results align with our findings presented in \figstext~\ref{fig:responsivesweep} and~\ref{fig:exposuresweep} where we observed sharp transitions between polarized and non-polarized outcomes as functions of tolerance and exposure.
The transition with respect to responsiveness was less pronounced, with low responsiveness yielding large variability in polarization outcomes.

\paragraph*{Multiple Ideological Dimensions.}

\begin{figure}[tbh]
    \centering
    \includegraphics[width=0.6\linewidth]{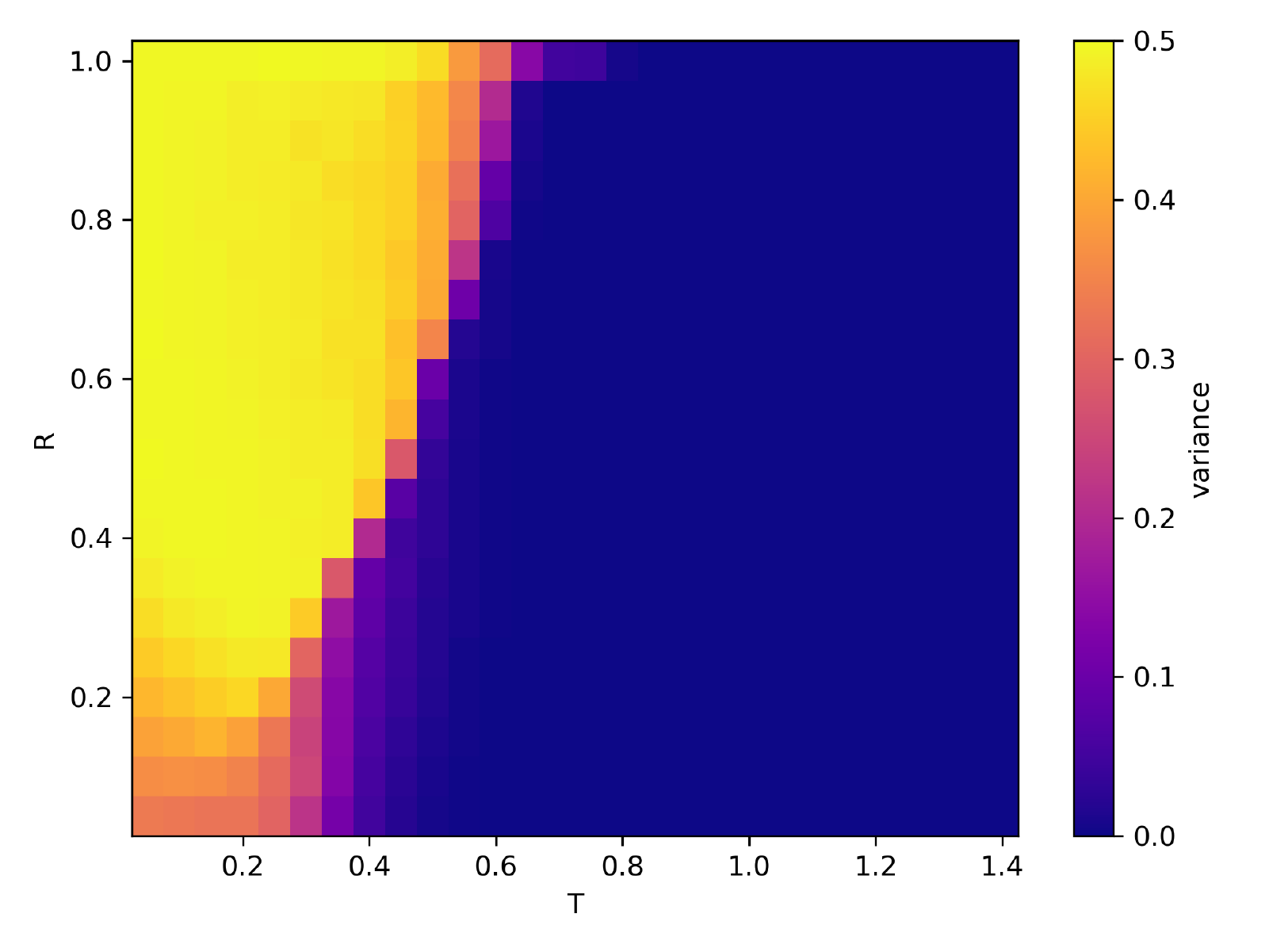}
    \caption{\textbf{The Effects of Tolerance (T) and Responsiveness (R) in Two Ideological Dimensions.}
    Average polarization of the population's ideological positions after 1,000,000 steps, averaged over 20 iterations for each $(T, R)$ pair.
    Tolerance is varied over $T = 0.05, 0.1, \ldots, 1.4$ while responsiveness is varied over $R = 0.05, 0.1, \ldots, 1.0$.
    As in one dimension (\figtext~\ref{fig:responsivesweep}), there is a phase change from extreme polarization (yellow) to convergence (dark blue) as $T$ increases.}
    \label{fig:responsivesweep2D}
\end{figure}

The effects of tolerance and responsiveness in two dimensions (\figtext~\ref{fig:responsivesweep2D}) are analogous to those in one dimension (\figtext~\ref{fig:responsivesweep}).
There is a phase change from extreme polarization with low tolerance ($T \leq 0.3$) to convergence with sufficiently high tolerance ($T \geq 0.7$) that is largely independent of responsiveness, $R$.
Repulsive extremists also appear in two dimensions, causing some runs along the phase boundary ($0.35 \leq T \leq 0.65$) to converge to intermediate polarization.

\begin{figure}[tbh]
    \centering
    \includegraphics[width=0.6\linewidth]{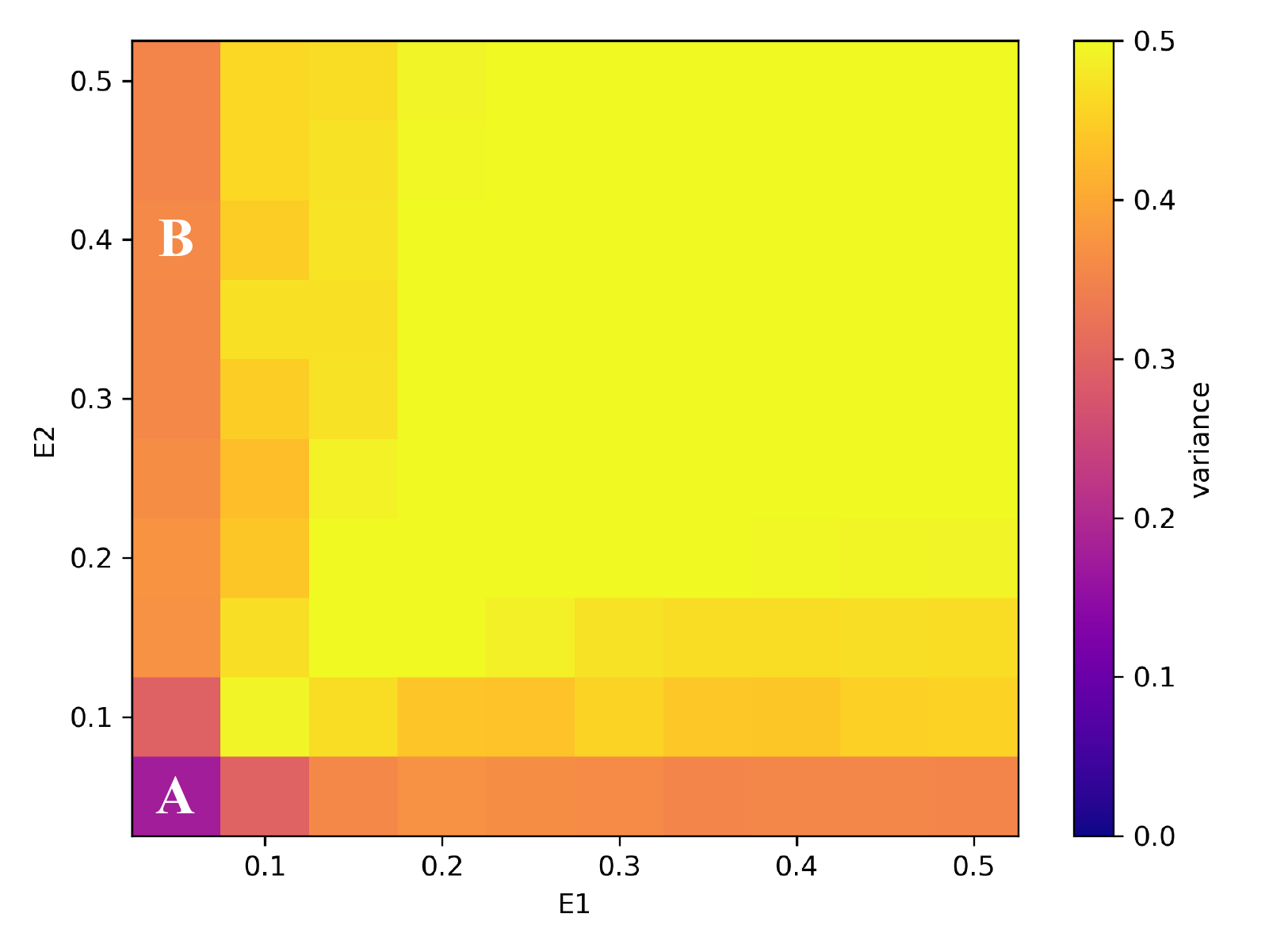}
    \caption{\textbf{The Effects of Exposure (E) in Two Ideological Dimensions.}
    Average polarization of the population's ideological positions after 2,000,000 steps, averaged over 20 iterations for each $(E_1, E_2)$ pair.
    Each ideological dimension's exposure is varied over $E_1, E_2 = 0.05, 0.1, \ldots, 0.5$.
    \textbf{(A)} and \textbf{(B)} indicate the $E_2 = 0.05$ and $E_2 = 0.4$ runs shown in \figtext~\ref{fig:exposureevo2D}.}
    \label{fig:exposuresweep2D}
\end{figure}

\figtext~\ref{fig:exposuresweep2D} shows the effects of different levels of exposure per ideological dimension.
Similar to the one-dimensional case, when $E_1, E_2 \geq 0.2$ actors often interact with and are repulsed by others beyond their tolerance, causing extreme polarization.
However, populations with low exposure even on just one dimension ($E_1 \leq 0.15$ or $E_2 \leq 0.15$) avoid extreme polarization for most degrees of exposure on the other dimension.

\paragraph*{Economic Self-Interest.}

\begin{figure}[tbh]
    \centering
    \includegraphics[width=0.6\linewidth]{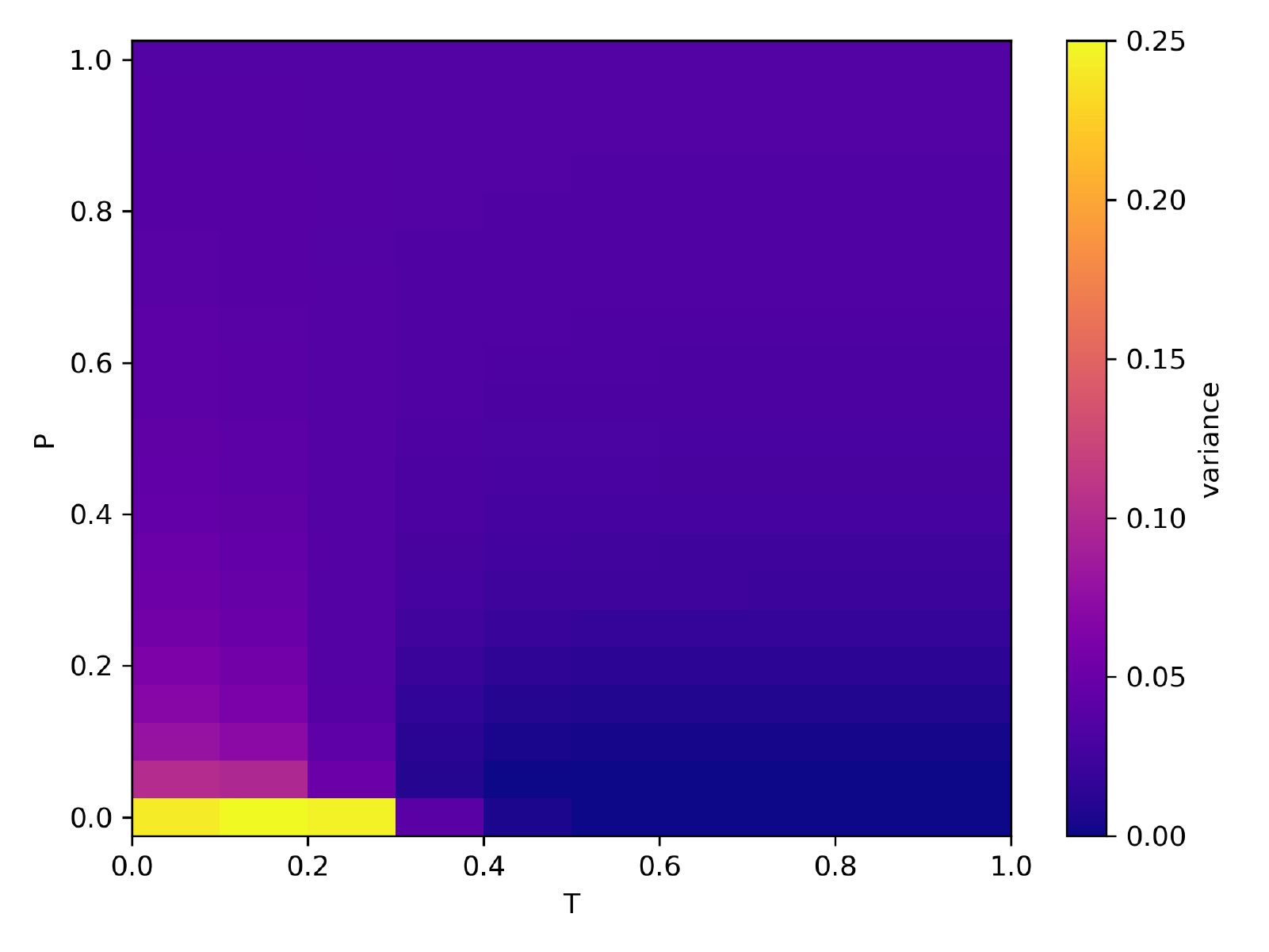}
    \caption{\textbf{The Effects of Economic Self-Interest (P) as a Function of Tolerance (T).}
    Average polarization of the population's ideological positions after 2,000,000 steps, averaged over 20 iterations for each $(T, P)$ pair.
    Tolerance is varied over $T = 0.05, 0.15, \ldots, 0.95$ and self-interest is varied over $P = 0\%, 5\%, \ldots, 100\%$.}
    \label{fig:selfinterestsweep}
\end{figure}

Parameter sweeps of tolerance $T$ and self-interest probability $P$ (\figtext~\ref{fig:selfinterestsweep}) confirm that even small levels of self-interest can greatly reduce polarization in populations that would have otherwise gone to the extremes ($T \leq 0.25$).
On the other hand, populations that would have otherwise converged ($T \geq 0.45$) no longer reach stable consensus because each actor continues to attract to its preferred position with probability $P$.
When self-interest is sufficiently high ($P \geq 40\%$) actors' attraction to their normally distributed preferred positions dominates the interaction effects entirely, yielding little variation from the initial opinion distribution.

\end{document}